\documentclass[fleqn,usenatbib,useAMS]{mnras}
\usepackage[table,xcdraw]{xcolor}
\usepackage{newtxtext,newtxmath}
\usepackage{graphicx}
\usepackage[utf8]{inputenc}
\usepackage[english]{babel}
\usepackage{subfigure}
\usepackage[T1]{fontenc}
\usepackage[T1]{fontenc}
\usepackage{hyperref}
\usepackage{rotating}
\usepackage{soul}
\usepackage[flushleft]{threeparttable}
\hypersetup{colorlinks,citecolor=blue,filecolor=blue,linkcolor=blue,urlcolor=blue}
\DeclareRobustCommand{\VAN}[3]{#2}
\let\VANthebibliography\thebibliography
\def\thebibliography{\DeclareRobustCommand{\VAN}[3]{##3}\VANthebibliography}

\usepackage{graphicx}	
\usepackage{amsmath}	

\title[AGN accretion disk sizes from ZTF survey]{Accretion Disk Sizes from Continuum Reverberation Mapping of AGN Selected from the ZTF Survey}
\author[Vivek Kumar Jha~et~al.]{
Vivek Kumar Jha$^{1,2}$,\thanks{E-mail: vivekjha.aries@gmail.com}
Ravi Joshi$^{3,4}$, 
Hum Chand$^{5}$,
Xue-Bing Wu$^{4,6}$,
Luis C. Ho$^{4,6}$, 
Shantanu Rastogi$^{2}$,
\newauthor
and Qinchun Ma$^{4,6}$ \\
\\
$^{1}$Aryabhatta Research Institute of observational sciencES, Nainital, \it {263002}; India\\
$^{2}$Deen Dayal Upadhyaya Gorakhpur University, Gorakhpur, \it {273009}; India\\
$^{3}$Indian Institute of Astrophysics, Koramangla, Bangalore,  \it{560034}; India\\
$^4$Kavli Institute for Astronomy and Astrophysics, Peking University, Beijing 100871, China\\
$^{5}$Central University of Himachal Pradesh, Dharamshala, \it{176215}; India\\
$^6$Department of Astronomy, School of Physics, Peking University, Beijing 100871, China}
\date{Accepted XXX. Received YYY; in original form ZZZ}
\pubyear{2021} 
\begin{document}
\label{firstpage}
\pagerange{\pageref{firstpage}--\pageref{lastpage}}
\maketitle

\begin{abstract}
 We present the accretion disk size estimates for a sample of  19  active galactic nuclei (AGN) using the optical $g$, $r$, and $i$ band light curves obtained from the Zwicky Transient Facility (ZTF) survey. All the AGN have reliable supermassive black hole (SMBH) mass estimates based on previous reverberation mapping measurements. The multi-band light curves are cross-correlated, and the reverberation lag is estimated using the Interpolated Cross-Correlation  Function (ICCF) method and the Bayesian method using the {\sc javelin} code. As expected from the disk reprocessing arguments, the $g-r$ band lags are shorter than the $g-i$ band lags for this sample. The interband lags for all, but 5 sources, are larger than the sizes predicted from the standard Shakura Sunyaev (SS) analytical model. We fit the light curves directly using a thin disk model implemented through the {\sc javelin} code to get the accretion disk sizes. The disk sizes obtained using this model are on an average 3.9 times larger than the prediction based on the SS disk model. We find a weak correlation between the disk sizes and the known physical parameters, namely, the luminosity and the SMBH mass. In the near future, a large sample of AGN covering broader ranges of luminosity and SMBH mass from large photometric surveys would be helpful in a better understanding of the structure and physics of the accretion disk.

\end{abstract}




\section{Introduction}

Accretion disks in AGN form as a result of inflowing matter onto the central Supermassive black hole [SMBH]~ \citep{1964ApJ...140..796S,
1969Natur.223..690L}. Various accretion disk models have been proposed ranging from the standard Shakura Sunyaev (SS) disk \citep{Shakura1973}, the slim disk \citep{abramowicz1988} to more complicated models like the advection dominated accretion disk presented by \citet{Narayan1995} and the relativistic accretion disk around a Kerr black hole presented by \citet{xin2005}. Based on these models, the accretion disk has a temperature profile, with higher temperatures in the inner regions and decreasing temperatures in the outer regions. The growing observational shreds of evidence suggest that the X-ray emission in the innermost regions of the AGN accretion disk are reprocessed as the UV/optical continuum flux. Further, the rapid X-ray variability and relative strength of the hard X-ray component point towards the presence of a corona in the proximity of the SMBH, which irradiates the accretion disk. However, the exact size and structure of this corona are not known precisely \citep[e.g., see][]{Meyer-Hofmeister2017, Arcodia2019, Sun2020}.


 According to the standard SS disk model, the effective
temperature $T$ at a particular location in the accretion disk $R_{disk}$ is related to the disk size and the SMBH mass as follows:
\begin{equation}
T (R_{disk})=\left(\frac{3GM_{BH}\dot{M}}{8\pi\sigma (R_{disk})^3}\right)^{1/4},
\label{eqn:basic}
\end{equation}
where $\sigma$ denotes the the Stefan-Boltzmann constant and 
$G$ and $\dot{M}$ represent the gravitational constant and the mass accretion rate respectively.  As is evident from this equation, the accretion disk size (R) scales inversely with the temperature (T) by a power law of index 4/3.

The photon wavelength ($\lambda_0$) at a particular disk radius is related to the temperature by Wein's law. The size of the accretion disk at the wavelength $\lambda_0$ can be determined by two observable parameters, the SMBH mass and the mass accretion rate. The radiation from the accretion disk is assumed to be composed of multiple black bodies, and the disk sizes at various wavelengths can be calculated using the equation below:

\begin{equation}
\centering
R_{\lambda}=R_{\lambda_0}\left[\left(\frac{\lambda}{\lambda_0}\right)^{\beta}\right]
\label{equation2}
\end{equation}

here,
$R_{\lambda_0}$ is the disk size at a reference wavelength $\lambda_0$ and $\beta$ is 4/3=1.33 for the standard SS disk. The temperature profile of the disk generated using this formalism can be used to test the application of the SS disk to the AGN population.  

Direct observation of these regions is difficult as they project to sub-micro arc second angular resolutions, which any current or near-future observational facilities cannot resolve. The disk sizes for a few AGN have been estimated using microlensing studies in \citep[see][]{Pooley2007, Dai2010, Morgan2010, Blackburne2011, Mosquera2013}. Another technique that remains promising is Reverberation Mapping (RM). Utilizing the variable nature of AGN at most time scales, RM substitutes spatial resolution with time resolution to measure the lag between the flux emitted from different regions, thereby enabling us to reconstruct the echo images that encode the information about the structure and kinematics of AGN innermost region \citep[see][]{Bahcall, Blandford1982, Peterson2006, Bentz2009}. Traditionally the RM technique was used to estimate the size of the Broad Line Region (BLR) using the cross-correlation between the accretion disk continuum and the H$\beta$ emission line,  which has yielded a significant correlation between the size of the BLR  and the AGN luminosity \citep{Kaspi2002, Bentz2009, Du2016, Du2018}. Resolving the accretion disk structure using RM through multi-band observations is possible assuming the {\it lamppost} model, which implies a disk being irradiated by X-rays originated in the innermost central regions, e.g., corona above the disk,  and the photons being reprocessed in the form of UV and optical wavelength emissions \citep{Cackett2007, Fausnaugh2017}. As the emission from the disk is expected to be of the black body type peaking at different wavelengths, depending on the temperature of the disk, continuous, simultaneous observations in multiple wavelengths ranges to cover hotter inner regions and cooler outer regions are expected to yield the structure and temperature profile of the accretion disk itself.


However, due to the complexity involved in executing such multi-band monitoring campaigns, accretion disk structure has been mapped for only a handful of AGN so far \citep[see][]{rosa2015, Fausnaugh2016, Starkey2017, Pei2017, Edelson2019, HernandezSantisteban2020}. Alternatively, with the availability of large all-sky surveys, it has been possible to constrain the accretion disk sizes based on interband lags for a sample of AGN using observations in the optical regime only. In \citet{Sergeev2014} the interband lags were detected for a sample of 14 AGN in the optical wavelengths. In \citet{Jiang2017}, inter-band lags based on the PanStarrs light curves were presented while \citet{Mudd2018} presented results for 15 quasars from the OzDES survey and \citet{Yu2018}  performed the analysis with around 23 quasars from the same survey but a different field. In \citet{Homayouni2020}, inter-band lags for around 33 quasars were calculated from the sample of SDSS RM quasars. These studies used the optical wavelength lags only, and the accretion disk profile from the X-ray (innermost regions) to optical wavelengths (outer accretion disk) could not be appropriately constrained. However, the presence of lags itself indicates evidence of the reprocessing of photons in the accretion disk. These results put an upper limit on the accretion disk sizes; nevertheless, the correlated reverberation disk signals are detected. Interestingly, although the interband lag scales with wavelength as a power law of index 4/3, the size of the disk has been observed to be 3-4 times larger than expected from the standard Shakura Sunyaev disk \citep{Fausnaugh2018}. Furthermore, \citet{Edelson2015} and \citet{Fausnaugh2016} have shown that the U band lags are significantly larger than the predictions from the SS disk models in some of the sources, and the Balmer emission from the BLR may likely affect the continuum flux.

The scaling of the physical parameters such as the SMBH mass, luminosity, and accretion rate with the size of the accretion disk has eluded us for the time being. Only a handful of AGN has both the accretion disk size measurements and SMBH mass estimation, making multi-band monitoring very important. In this work, using the publicly available Zwicky Transient Facility (ZTF)\footnote{\url{https://www.ztf.caltech.edu/}}, we aim to constrain the accretion disk sizes and their correlations with the physical parameters such as the SMBH mass and bolometric luminosities of  AGNs with well-constrained SMBH masses through prior RM studies. This paper is structured as follows: In Section \ref{section2} we present the sample of AGN studied in this work along with the details of the observations and data. In Section \ref{section4} we present the analysis based on various lag estimation methods, and the results of our comparison of disk sizes with the standard SS disk, followed by the correlations with various physical parameters, are presented in Section \ref{section5}. The discussion is presented in Section \ref{section6} and we conclude our results in Section \ref{section7}.

\begin{figure*}

\includegraphics[width=17cm,height=8cm]{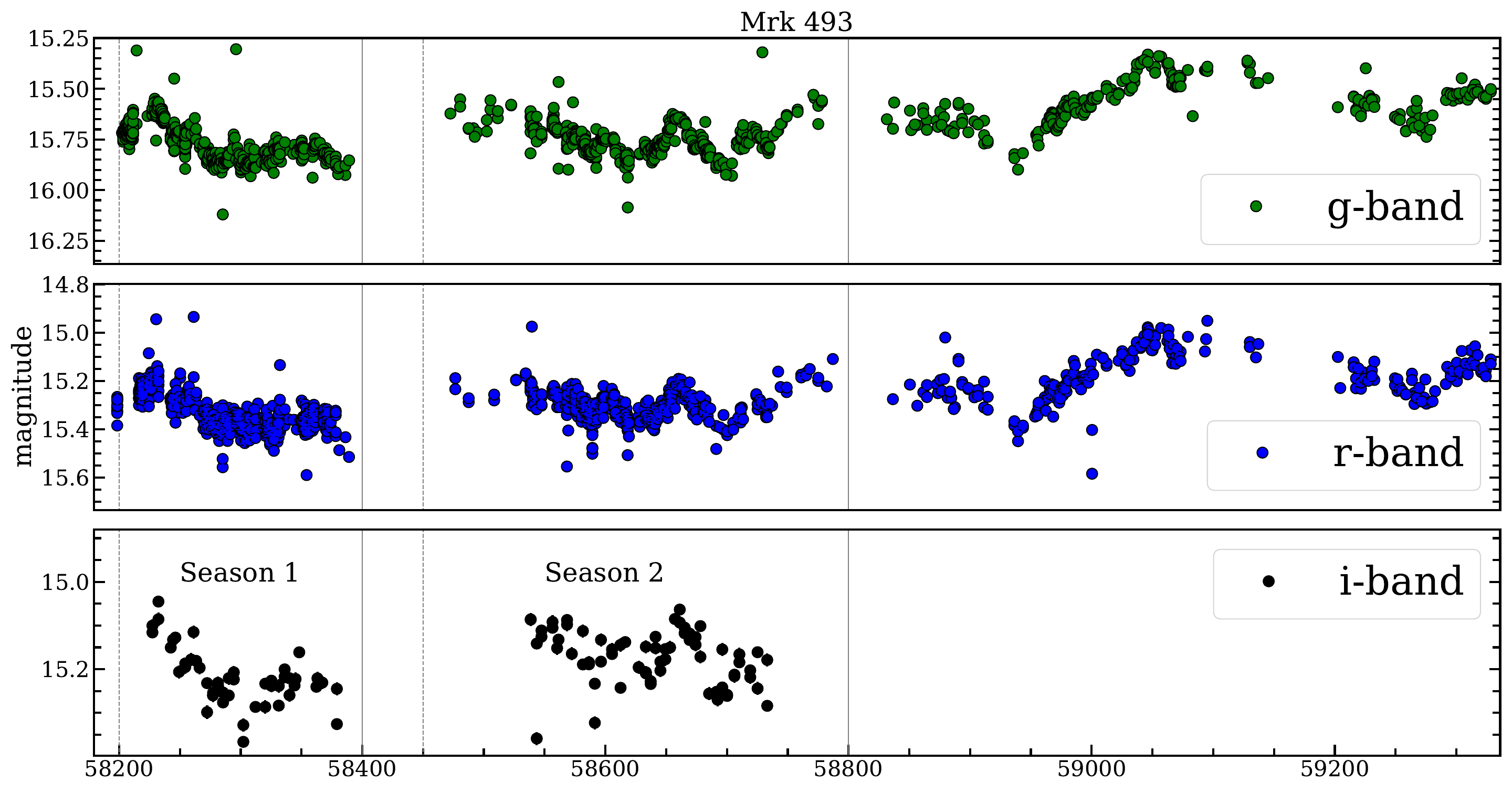}

\caption{ The $g$, $r$, and $i$-band light curves for Mrk  493. There are seasonal gaps between the ZTF light curves after MJD 58400 and MJD 58800, based on which the light curves are divided in season 1 (MJD ranging from 58200 to 58400) and season 2 (MJD ranging from 58450 to MJD 58800), respectively. The dashed lines denote the beginning of a season, followed by a solid line denoting the end of a season. The $i$-band observations are not available beyond MJD 58850. }
\label{Figure0}
\end{figure*}



\begin{figure*}

\subfigure{\includegraphics[width=8.5cm,height=5cm]{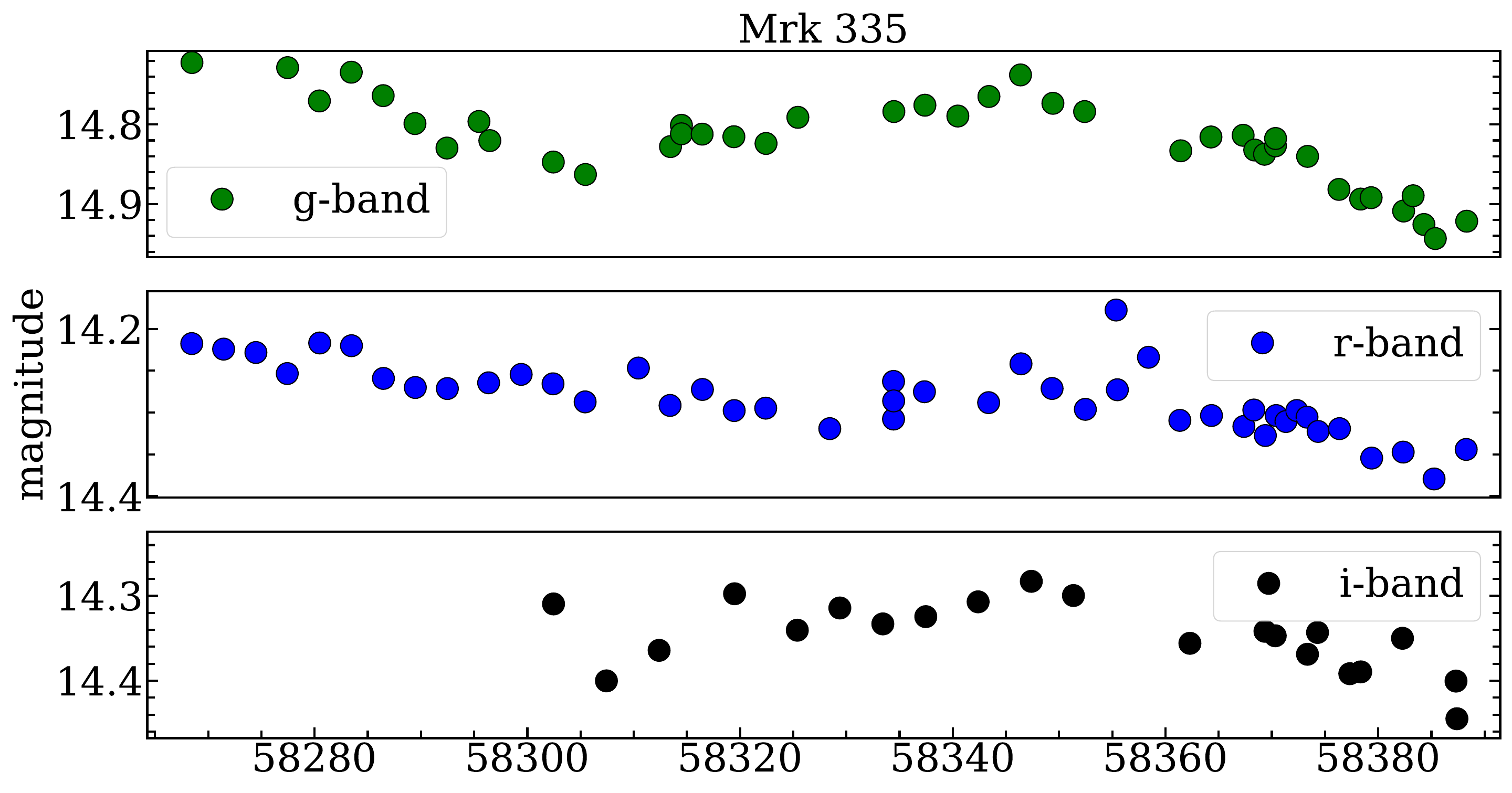}}
\subfigure{\includegraphics[width=8.5cm,height=5cm]{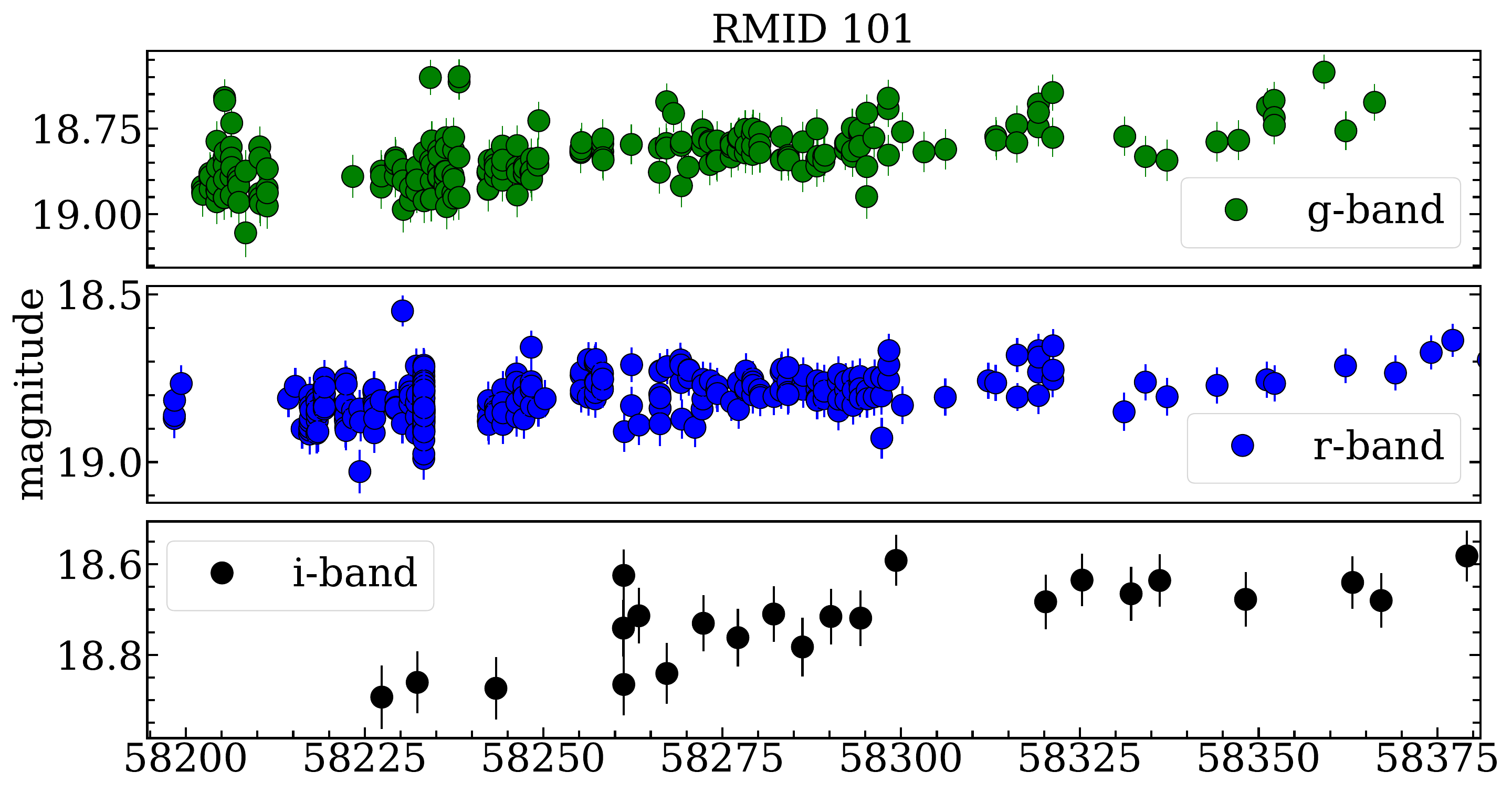}}
\subfigure{\includegraphics[width=8.5cm,height=5cm]{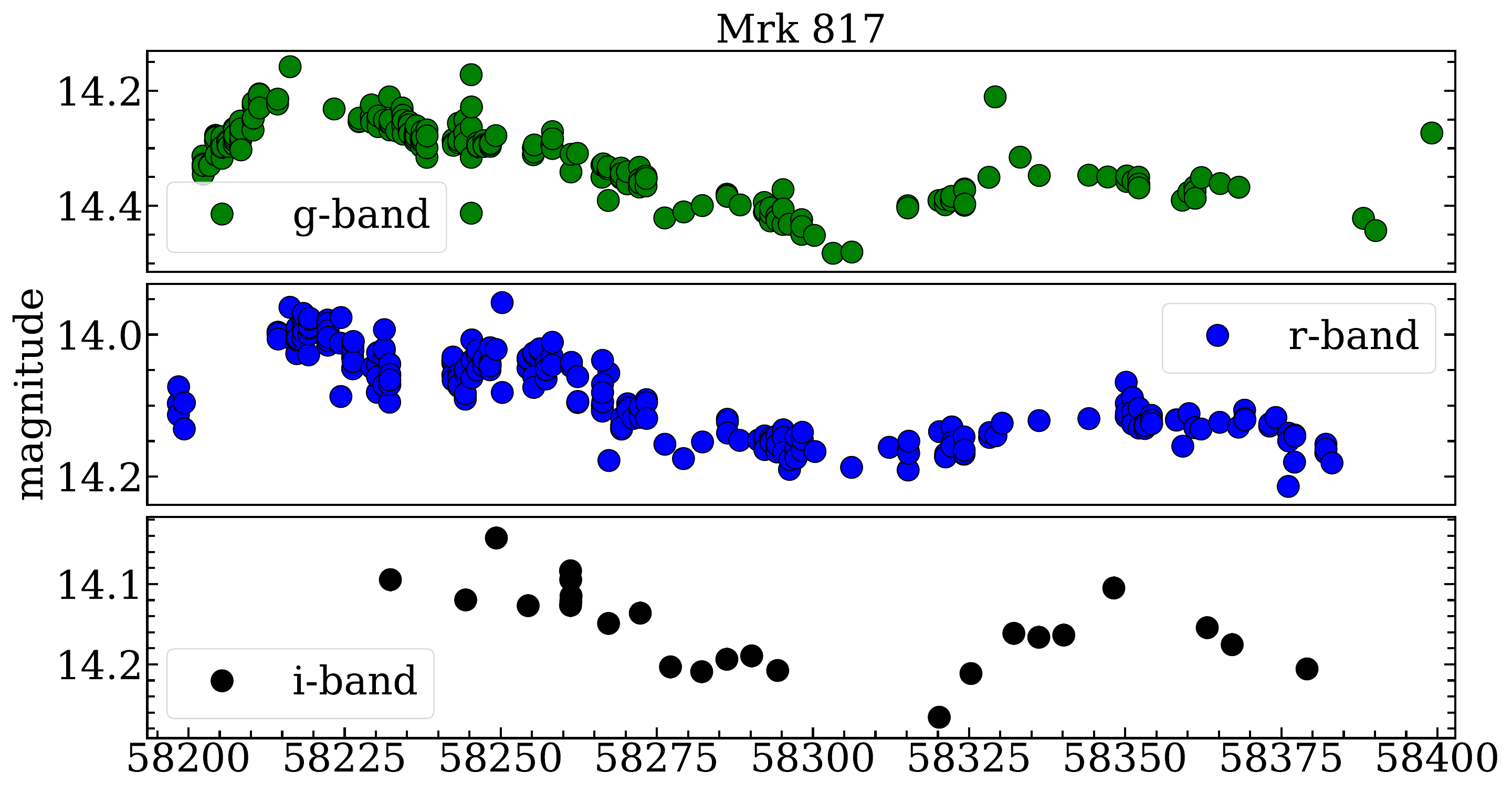}}
\subfigure{\includegraphics[width=8.5cm,height=5cm]{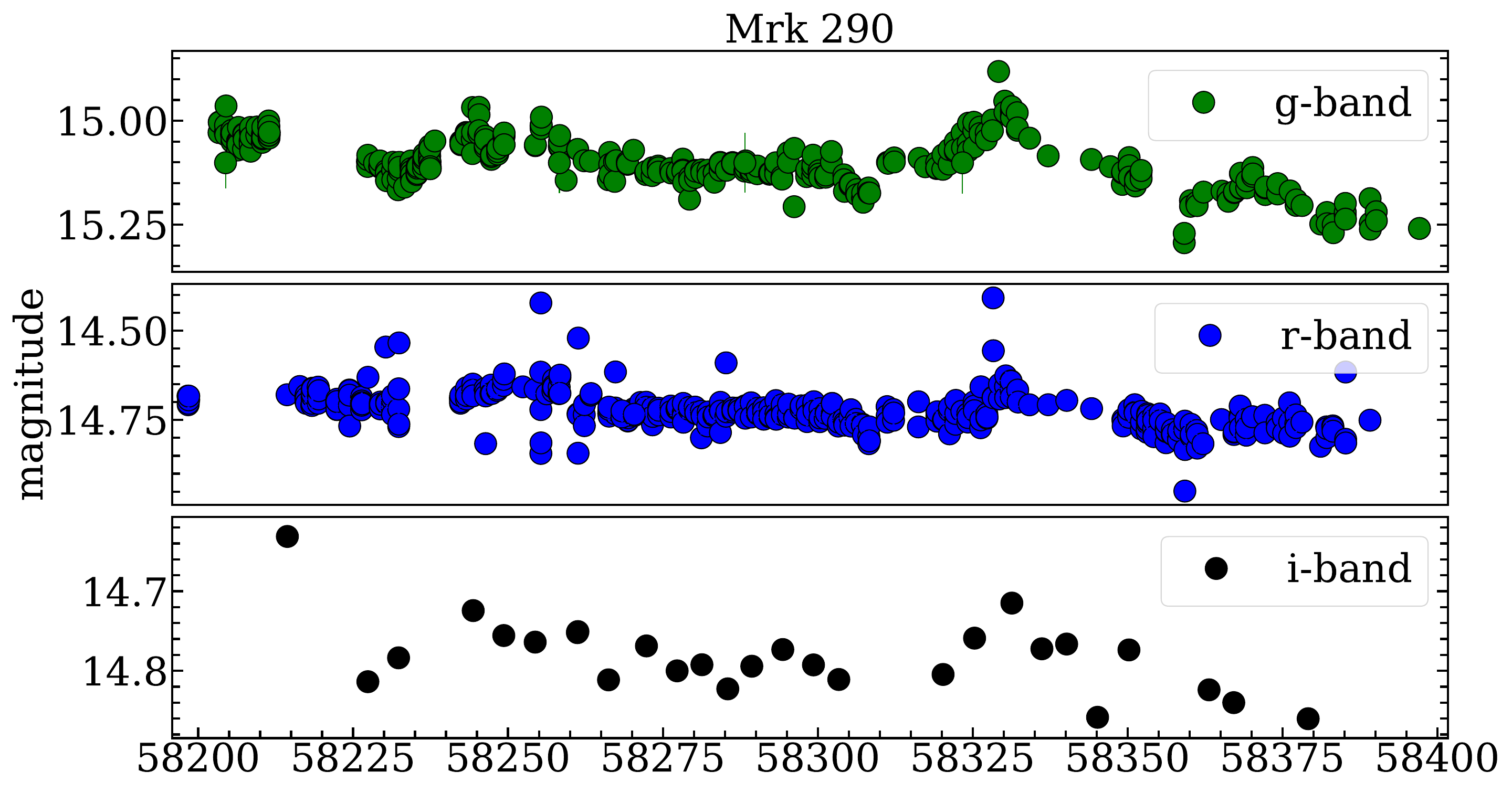}}
\subfigure{\includegraphics[width=8.5cm,height=5cm]{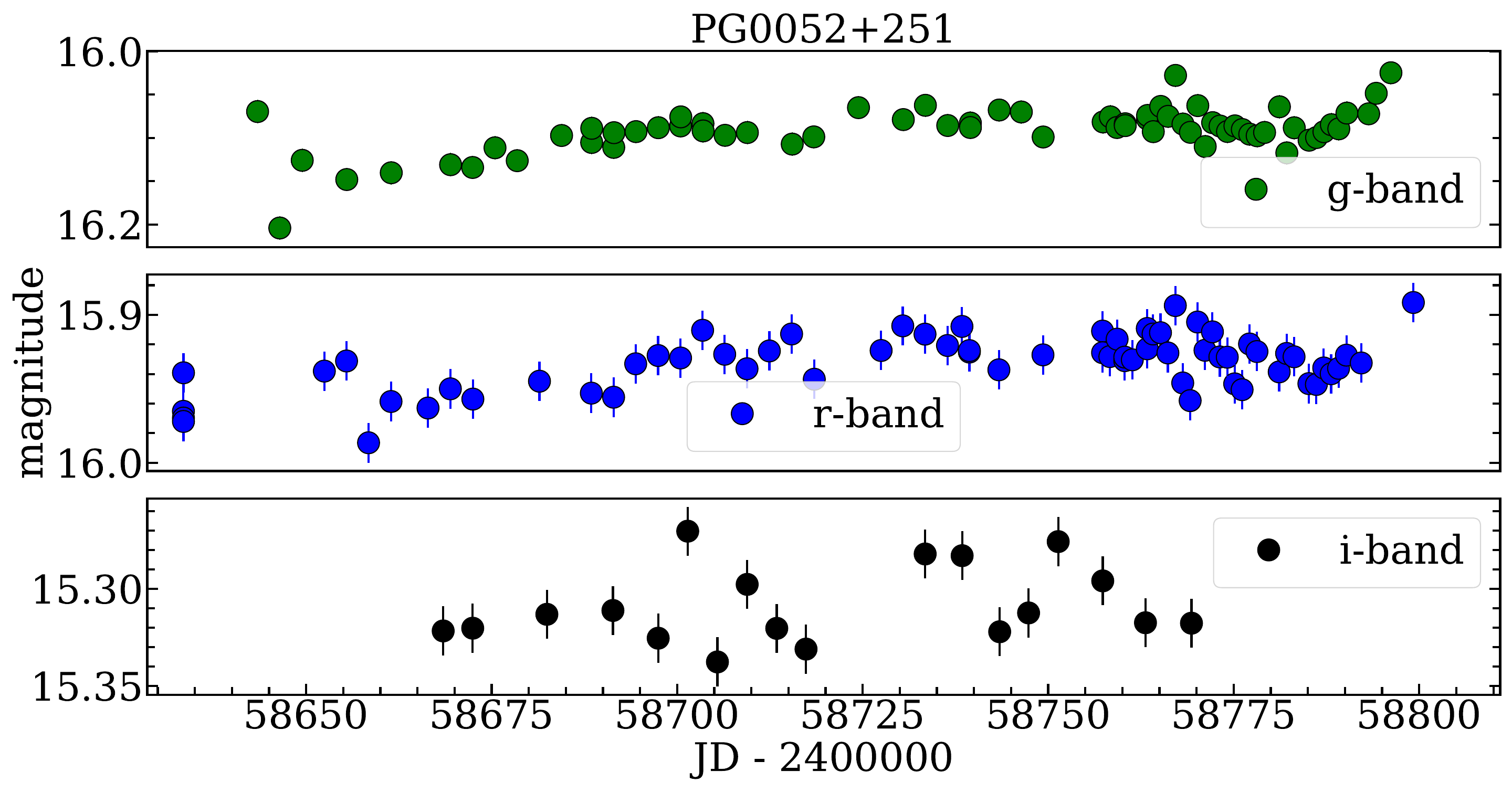}}
\subfigure{\includegraphics[width=8.5cm,height=5cm]{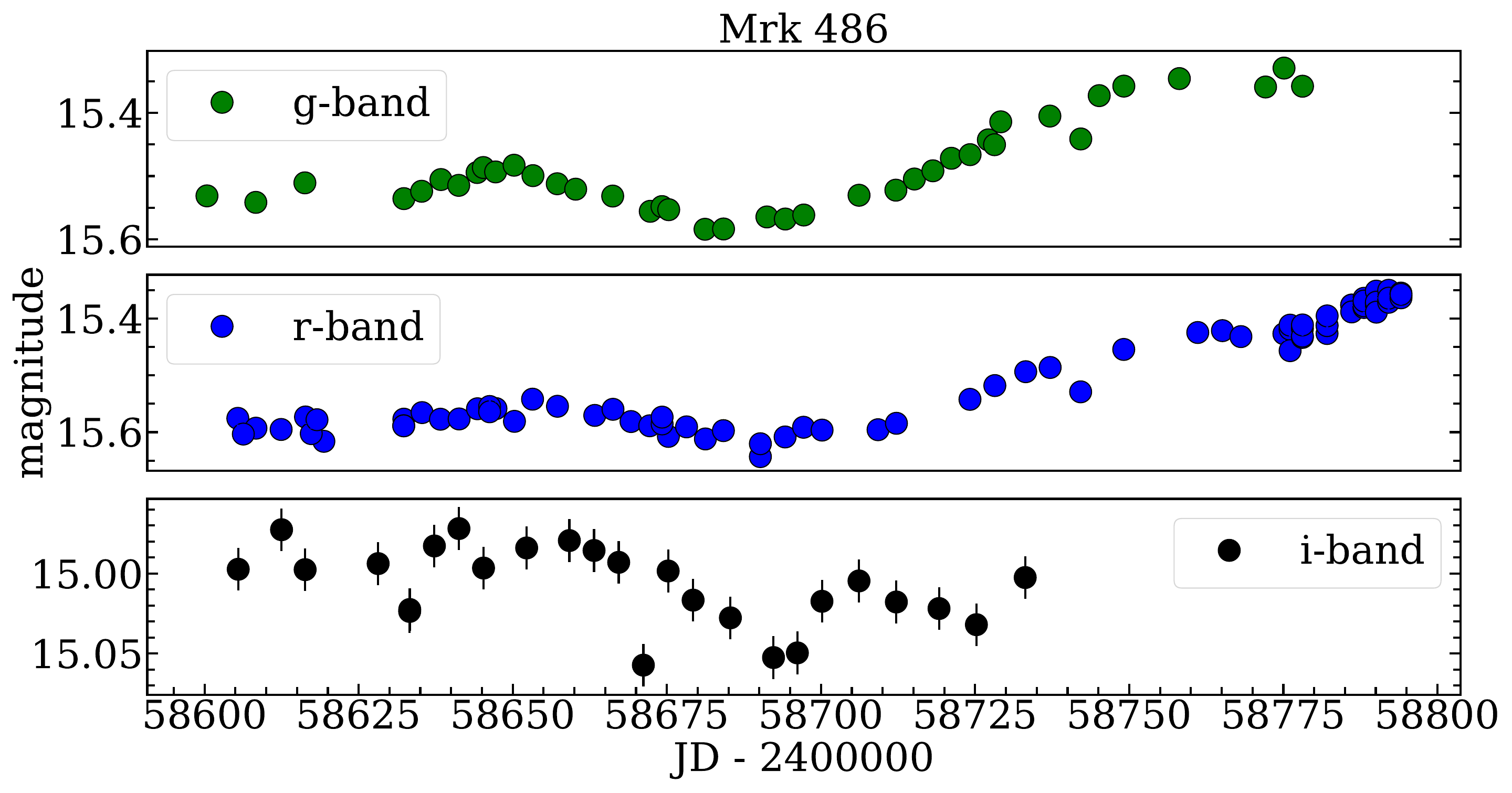}}

\caption{The  $g-$band (green), $r-$band (blue), and $i-$band (black) light curves for Mrk 335 and RMID101 are shown in the top panel, for Mrk  817 and Mrk  290 are shown in the middle panel, and for PG0052+251 and Mrk 486 are shown in the bottom panel. The first four light curves are from season 1, while for the sources in the bottom panel, light curves are from season 2. The units for magnitude are calibrated magnitudes as available from the ZTF database. The light curves include the error bars; however, the error bars are smaller than the markers used here for some of the sources. The statistics for these light curves are available in Table \ref{Table2}.}
\label{Figure1}
\end{figure*}

\begin{figure*}
\subfigure[Mrk  486]{\includegraphics[width=5cm,height=4cm]{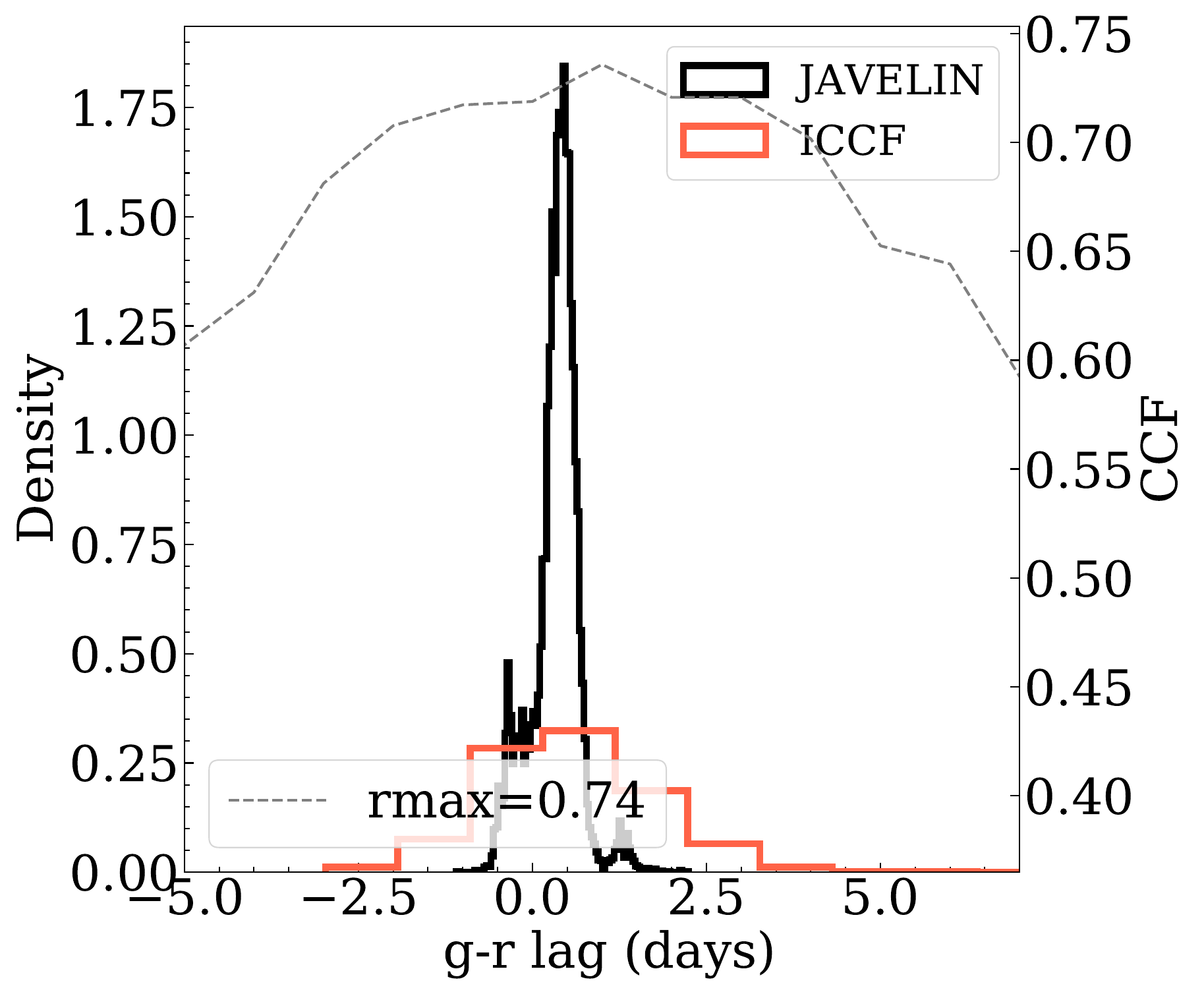}}
\subfigure[Mrk  335]{\includegraphics[width=5cm,height=4cm]{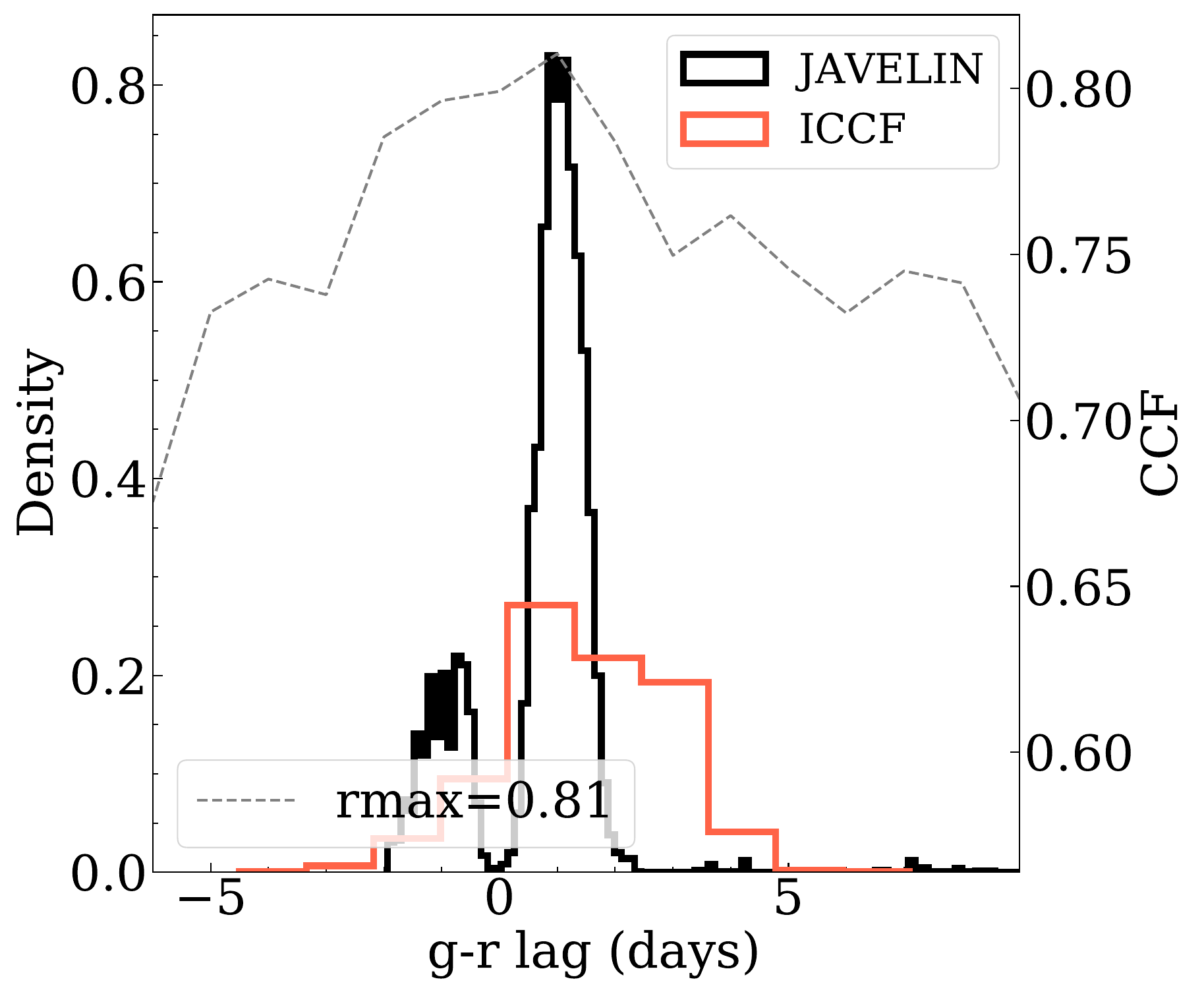}}
\subfigure[1RXS J1858+4850]{\includegraphics[width=5cm,height=4cm]{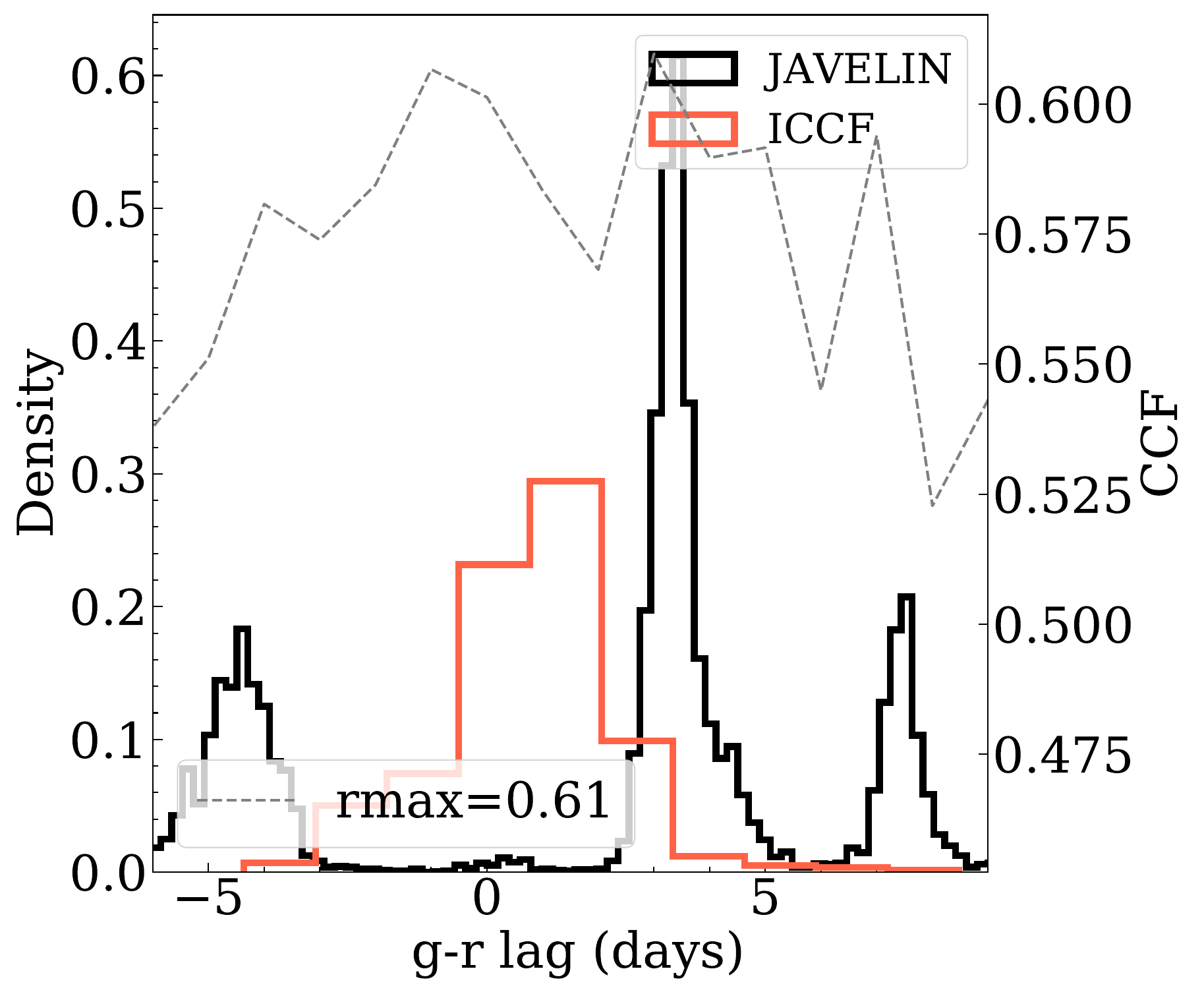}}
\subfigure{\includegraphics[width=5cm,height=4cm]{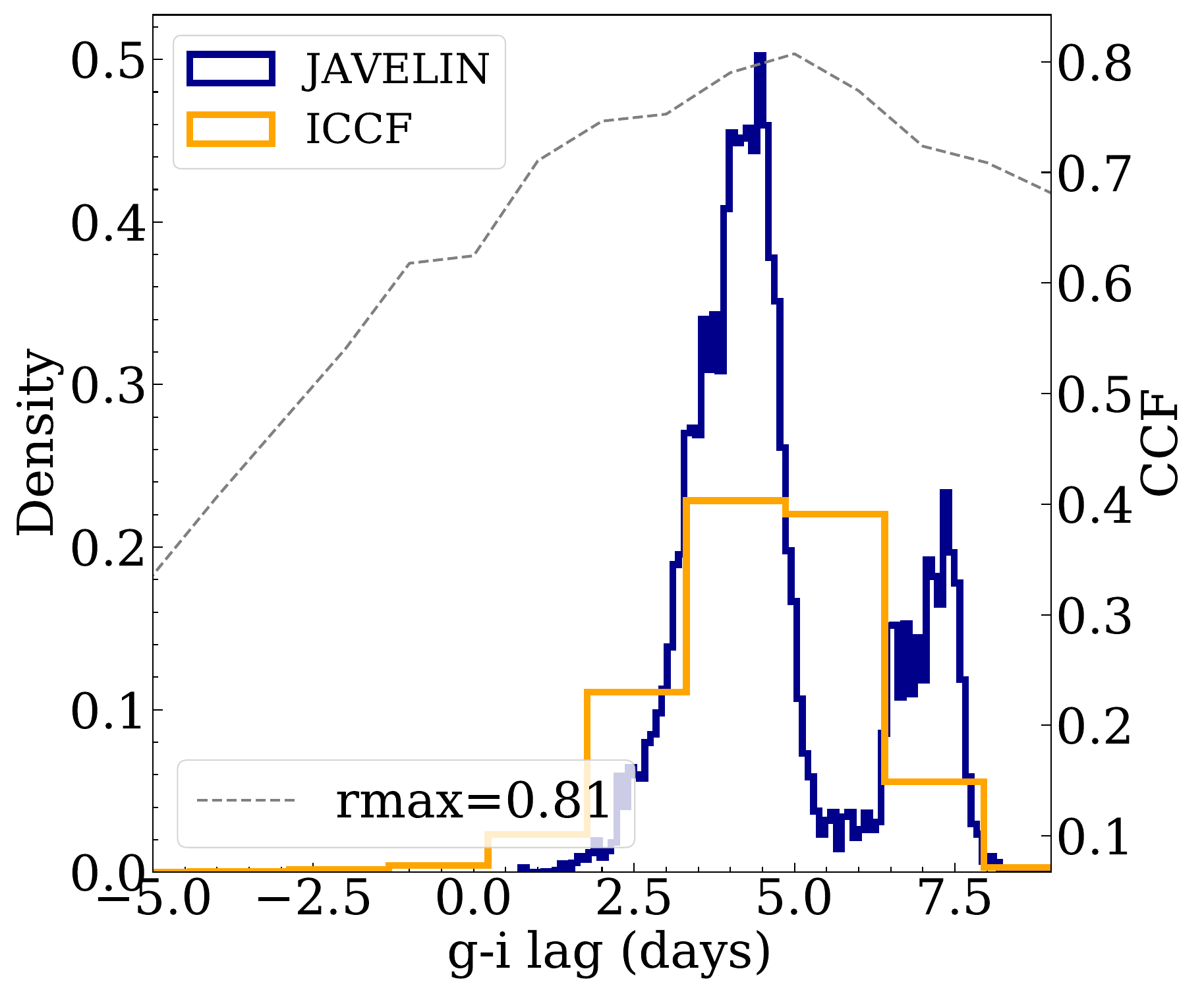}}
\subfigure{\includegraphics[width=5cm,height=4cm]{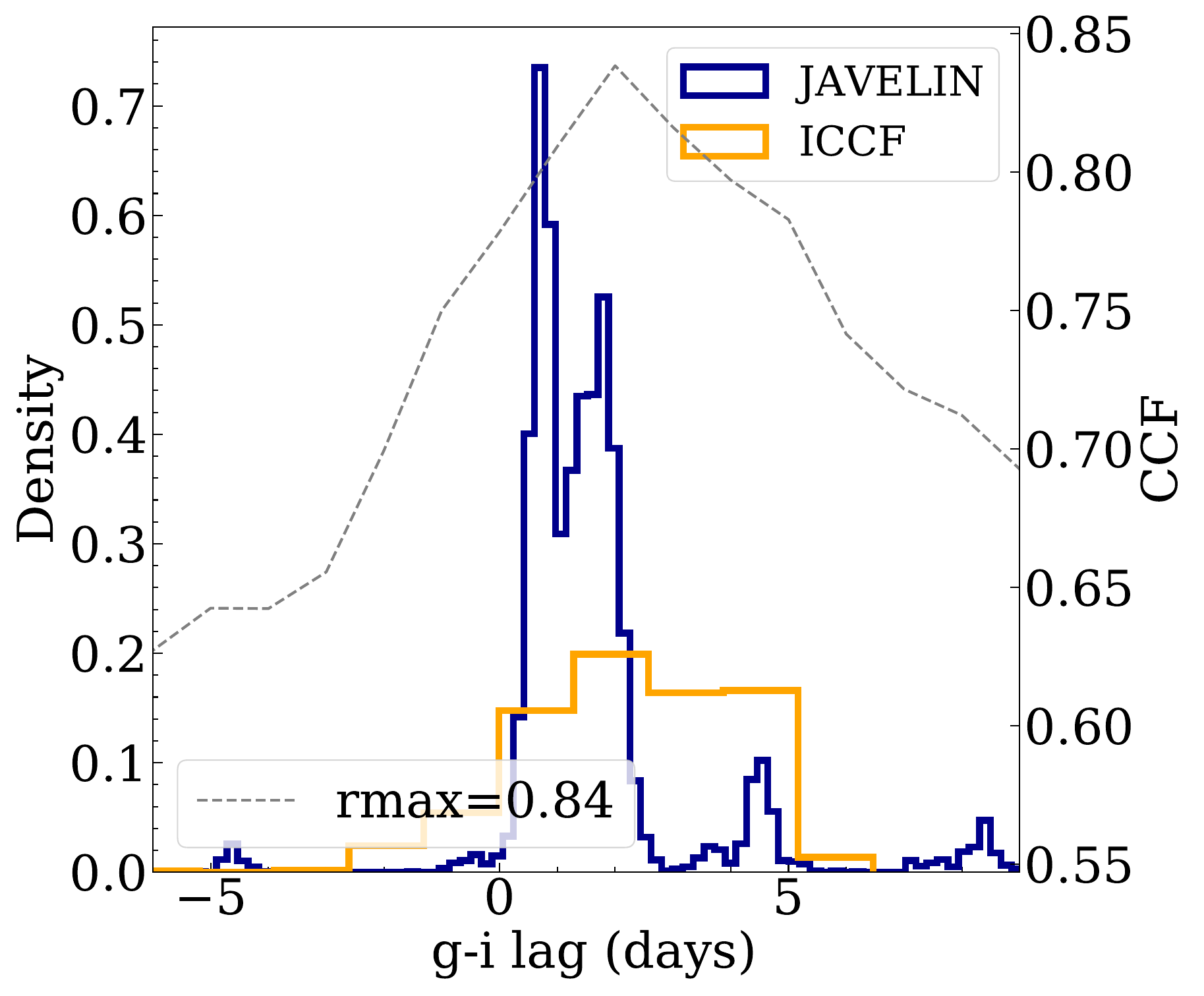}}
\subfigure{\includegraphics[width=5cm,height=4cm]{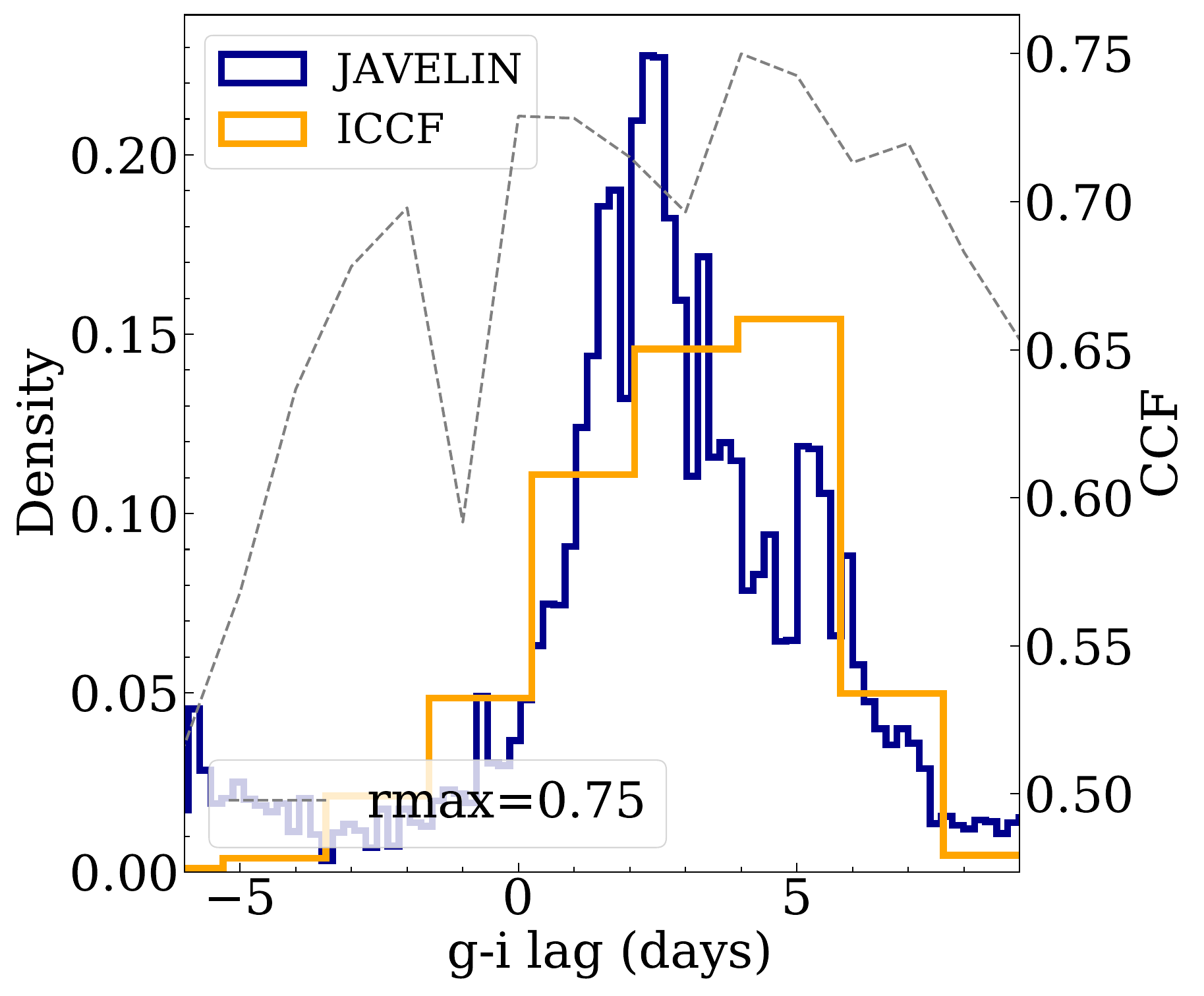}}

\caption{ The $g-r$ lag distribution obtained using {\sc javelin}  (black) and ICCF (red) for 3 AGN: Mrk  486 (left), Mrk  335 (middle) and 1RXSJ1858+4850 (right) are shown in top panel. The $g-i$ lag distribution is shown in the bottom panel with the {\sc javelin} distribution denoted as blue color and the ICCF distribution denoted as orange color. The CCF obtained using the ICCF method is shown in grey color on the plots. The maximum correlation coefficient obtained using the CCF method is denoted as rmax. Table \ref{Table3} contains the results obtained from the ICCF and the {\sc javelin}  methods.}
\label{Figure2}
\end{figure*}

\section{The sample and data}
\label{section2}

To measure the accretion disk sizes and thereby test the accretion disk models, we assembled the reverberation mapped AGN selected from various RM campaigns, including 61 AGNs from the AGN Black hole mass database\footnote{\url{http://www.astro.gsu.edu/AGNmass/}}\citep[see][]{Bentz2015}, 44 AGNs from Sloan Digital Sky Survey - Reverberation Mapping (SDSS RM) project \citep{Grier2019}, and 25 AGNs from the Super Eddington Accreting Massive Black Holes (SEAMBH) sample available in \citet{Du2015, Du2016, Du2018}. However, 8 AGNs from SEAMBH list were already included in the AGN mass database. Thus, we were left with a total of 122 sources. Having an advantage of accurate BH mass estimation, an essential input for the disk size measurement, the sample spans a significant range of luminosities (42.57$\leq$ log(L$_{5100\AA}$) $\leq$ 44.97) and redshifts (0.013 $\leq$ z $\leq$ 0.646) providing a homogeneous set to study the AGN population.

For the multi-band continuum light curves of these AGNs, we explore the Zwicky Transient Facility (ZTF) time-domain survey \citep{Graham2019}. The ZTF uses the 48 inch Samuel Oschin Schmidt telescope with a field of view of 47 deg$^2$ to map the sky in $g$, $r$, and $i$ optical bands with a typical exposure time of 30 seconds, reaching a magnitude $\sim$20.5 in r$-$band \citep{Bellm2019}. The processing of the data is done using the Infrared Processing and Analysis Centre (IPAC) pipeline \citep{Masci2019}. In addition, the average cadence of 3 days makes it a comparatively better survey for continuum RM observations which has otherwise been not possible in some recent works using all-sky surveys \citep[see][]{Jiang2017, Mudd2018, Yu2018}. We obtained the light curves using the  ZTF-API \footnote{\url{https://irsa.ipac.caltech.edu/docs/program_interface/ztf_lightcurve_api.html}} by providing the positions of individual objects in terms of their  RA, DEC sky positions. Observations for all but 5 AGNs were available for at least one epoch in each band. We note that the $g$ and $r$ bands are well sampled in ZTF while the $i-$ band has the least number of observations (see Figure \ref{Figure0}). Keeping this into account, we set a criterion of at least 15 observation epochs in each band. This limit was kept in order to obtain enough reverberation signal in order to get the interband lags. Since there are gaps between the observations in all the three bands, we divide the light curves in {\it two seasons}, the first season (season 1) running between MJD 58200 to 58400 broadly and the second season (season 2) running between 58450 and 58800. Since the light curves are sampled well, and the lags we expect are of the order of a few days, this division will not affect the lag estimates, and dividing the light curves has an additional advantage of getting two disk size measurements between these intervals.

This criterion resulted in the final sample of 57 reverberation mapped  AGNs for season 1, including 34 sources from the SDSS-RM project, 23 AGNs from the AGN-Mass sample, and 2 AGNs from the SEAMBH sample, which are part of the AGN-Mass sample. In season 2, our sample size was limited to 22 AGN primarily because of the poor sampling in the $i$- band. Light curves for some of the sources are presented in Figure \ref{Figure1}. We moved forward with the final set of 57 AGNs from season 1 and 22 AGN from season 2 for inter-band lag estimation. All the 22 sources available in season 2 also satisfied the criterion in season 1 and thus, were part of the 57 sources selected from season 1. This provides us with the opportunity to measure the disk sizes for 22 AGN in two different time ranges.

\section{Measurement of inter band lags}
\label{section4}

Before measuring the lags between the multi-band continuum light curves, we first exclude any possible outliers in the light curves by applying a 3$\sigma$ clipping. We employ two most commonly used lag estimation methods, namely the Interpolated Cross-Correlation  Function (ICCF) method \citep{Peterson1998} and the {\sc javelin}  method \citep{Zu2011} which employs a Bayesian approach to estimate the lags. Given that the expected lags between the $g$ versus $r$ band are expected to be shorter than the $g$ versus $i$  band, we used the  $g$-band light curves as a reference for estimating the interband lags.

\subsection{{\sc javelin}  based lags}

We use the publicly available code {\sc javelin} \footnote{\url{https://github.com/nye17/JAVELIN }} to estimate the interband reverberation lags. It models the variability of the AGN as a Damped Random Walk (DRW) or, in other terms, an Ornstein Uhlenbeck (OU) process for timescales longer than a few days \citep{Zu2011, Zu2013a}. This approach has been demonstrated by \citet{Kelly2009} for modeling 100 quasar light curves from the OGLE database and further by \citet{MacLeod2010} for a sample of around 9000 quasars from the SDSS Stripe 82 region. The covariance function for the DRW process takes the following form:

\begin{equation}
S(\triangle t)= \sigma_{\tiny d}^2 e^{(-|\triangle t/ \tau_{d}|)}
\end{equation}

 Where $\triangle t$ is the time interval between the two epochs and $\sigma_{d}$ and $\tau_{d}$ are the amplitude of variability and the damping time scale, respectively. Lag estimation in {\sc javelin} is based on the assumption that the responding light curve variability is a scaled, smoothed, and displaced version of the driving light curve. Here, we took the  $g$-band light curve as the driving light curve, and the $r$ and $i$ band light curves were taken as the responding light curves while estimating the lags. Initially, using the driving light curves, we build the model to determine the parameters $\sigma_{d}$ and $\tau_{d}$. Then we fit the other two band light curves using this model to build the distribution for the time lag, the top hat smoothing factor, and the flux scaling factor. 
 
We derive the lag, tophat width, and scale factor distribution by shifting, smoothing, and scaling the light curves. We use  20,000 Markov Chain Monte Carlo (MCMC) chains to get the best fit parameter distribution. We set the lag limits to [-50, 50] days for the initial run. Given that the observed inter-band lags for various AGNs in previous studies were typically smaller than 10 days \citep[see,][]{Jiang2017, Mudd2018, Yu2018, Homayouni2020}, we rejected the sources for which the lags were obtained outside the limits of [$-$10,10] days as these estimates could be unphysical.


Recall that in season 1, we had 57 sources with sufficient sampling (no. of observations in $i$  band $\geq$ 15), and in season 2, we had 22 sources with sufficient sampling. Out of 57 sources in season 1, we could constrain the lag within the range of [$-$10, 10] days using {\sc javelin}  for 25 sources, while in season 2, out of 22 sources, we could constrain the lags for 19 sources within the same limits. Comparing the lag estimates between the two seasons, we find that lags for 14 sources were available in both season 1 and season 2; hence we have two measurements for this sample of sources while we have 11 individual measurements from season 1 and 5 individual measurements from season 2.

\subsection{ICCF based lags}

The Interpolated Cross-Correlation  Function (ICCF) method \citep{Gaskell1987, Peterson1998, Peterson2006} is quite frequently used to estimate inter-band reverberation lags in reverberation mapping studies. We used the  $g$-band light curve as the model light curve, which was assumed to drive the variations in the other two light curves. The observational gaps were interpolated linearly to create a uniformly sampled dataset, and the driving and responding light curves were sampled together, taking the effect of uncertainties into account. These interpolated light curves were cross-correlated to calculate the Cross-Correlation Function (CCF). This step was repeated multiple times to build the Cross-correlation Centroid distribution (CCCD) and Cross-correlation Peak distribution (CCPD). The CCCD has been used frequently as indicative of the interband lag \citep[e.g., see,][]{Peterson2006, Homayouni2020}. The interpolation frequency was set to 0.5 days as the ZTF light curves are quite well sampled in the $g$ and $r$ bands. The lag uncertainties were measured by employing the flux randomization (FR) and the random subset selection (RSS) sampling method over 5000 iterations with a significance of $r_{max} \leq 0.5$. We used a {\sc python} implementation of this method, known as PyCCF developed by \citet{Sun_pyccf}.

The $g-r$ lags were shorter than the $g-i$ lags for most of the cases. For instance, in the case of Mrk  335, we obtained a lag of $1.5^{+1.2}_{-1.3}$  days between the $g$ and $r$ bands while we obtained a lag of $2.4^{+1.6}_{-2.0}$ days between the $g$ and $i$ bands.  We note that ICCF lags suffer significant uncertainties compared to other methods \cite[][]{Fausnaugh2017, Li2019, Yu2018, Homayouni2020} which is likely because the ICCF method linearly interpolates between the data points, rather than assuming an underlying model when there are observational gaps. We estimated the lags using the ICCF  method for only the sources for which {\sc javelin}  lags were reasonable. The reason was that since {\sc javelin}  directly models the light curves as compared to the linear interpolation applied in the ICCF method, we expect the lag estimates to be more robust in the case of {\sc javelin}  as has been observed using simulations \citep{Li2019}. However, the lags estimated using the ICCF method serve as a check for the lags estimated through {\sc javelin}. Figure \ref{Figure2} shows the distribution of lags obtained between the $g-r$ and $g-i$ bands for a few of the sources in our sample using both methods. The estimated lags are listed in columns B, C, D, and E of Table \ref{Table3}. Out of the sources with lag estimates in the reasonable range, we picked up the sources where at least one method showed positive lags that increase with the wavelength in any of the seasons. This criterion resulted in the final sample of 19 sources which are the sources we use for further analysis (information available in Tables \ref{Table1}, \ref{Table2}, \ref{Table3}). The lags obtained were larger than that predicted by the Standard SS disk model for 14 sources in this sample, while for 5 sources, the lags were within the expectations of the SS disk models.
 
We note that for 4 AGN in our sample, previous interband continuum lags are available in the literature, which can be directly compared with our lag measurements. For Mrk  335, in \citet{Sergeev2014}, 
the B-R lag is reported to be $2.36^{+0.74}_{-0.85}$ days, and the B-I lag is reported to be $2.33^{+0.30}_{-1.86}$ days. We obtained a $g-r$ band lag of $1.5^{+1.2}_{-1.3}$ days and a $g-i$ band lag of  $2.4^{+1.6}_{-2.0}$ days. Even though the filter set in ZTF is a bit different than the filters used by \citet{Sergeev2014},  the interband lags between the two can be compared as the two sets of filters cover similar wavelength ranges. Mrk  817 is being monitored as part of the AGN STORM-II campaign \citep{Kara2021} and based on the initial results, the $g-r$ band lag is 
$1.50^{+0.86}_{-0.76}$ days and the $g-i$ band lag is $2.31^{+0.70}_{-0.86}$ using the ICCF method, while we obtain a lag of 
$2.5^{+1.5}_{-1.5}$ days between the $g-r$ band and a lag of $3.6^{+1.1}_{-1.4}$ days between the $g-i$ band using the same methods. These results are in agreement within the limits of the uncertainties. Furthermore, the interband lags between the $g$ and $i$ bands were obtained for RMID 101 and RMID 300 in \citet{Homayouni2020} as $1.54^{+2.06}_{-3.08}$ and $2.93^{+1.06}_{-4.24}$ days respectively using the ICCF method. We obtained a lag of $1.5^{+5.5}_{-5}$ and $2.0^{+6.5}_{-4.5}$ days respectively for these sources using the ICCF method.

Our $g-r$ band lags have relatively smaller uncertainties than $g-i$ band lags, likely due to inadequate sampling in the $i$  band in the ZTF survey. To address the impact of limited cadence and the uncertainties in light curves, first, we took the g-band light curves for Mrk 335, which is at lower redshift and has smaller uncertainties in the light curve, and SDSS RMID101, which is at higher redshift and having relatively larger uncertainties in the light curve. We introduced lags ranging from 0.5 to 3 days with an interval of 0.5 days by shifting the g-band light curve accordingly. We could recover lags up to 1 day for Mrk 335 with typical uncertainties of $\pm$ 0.2. However, in the case of RMID101, where the uncertainty in the light curves is larger, we could recover lags larger than 1.5 days within the uncertainties of about $\pm 0.5$. Next, to check the effect of the length of light curves in the recovery of lags, in order to mimic the lag recovery using i-band light curves, we reduced the number of data points in the responding light curves of Mrk 335 to 15 data points (the minimum number of points used in this study) by randomly sampling the light curve and re-estimated the lags using JAVELIN. In this case, we could recover the lags beyond 1.5 days with minimal deviation, but faced greater uncertainties of $\pm$ 1 day. Based on the above exercise, we note that the uncertainties in the data points and the length of light curves introduces large uncertainties in recovering the inter-band lags.

Another concern is that emission from the BLR emission lines may contribute to the accretion disk continuum. However, the variability timescale of BLR is known to be higher than the timescales for disk emission based on the reverberation mapping measurements of the $H\beta$ emission line. Further, recent works \citep[e.g.][]{Jiang2017} have concluded that the accretion disk continuum contributes more than 90 \% flux to the total flux, and thus emission line contamination may not be a significant contribution to broadband variations. However, the diffuse continuum may affect the lag estimates, especially for sources where the excess lags in the bands nearer to the UV wavelength have been reported \citep{Edelson2015}. In  \citet{korista2019} the effect of the diffuse continuum (DC) has been found to affect the delays w.r.t the 1158 \AA{} band. In our sample, 4 sources have redshifts where the DC can contribute in the g-band. However, in \citet{Fausnaugh2016}, detailed simulations have yielded that the interband continuum lags can be biased by a factor of 0.6$-$1.2 days due to BLR contamination. Based on this, we conclude that the uncertainties encountered in the lag estimates for these sources may account for this offset.\\

\begin{table}
\caption{Physical properties of the sample of reverberation mapped AGNs used in this study.}
\fontsize{6}{8.0}\selectfont
\label{Table1}
\begin{tabular}{lccccc}
\hline
\hline
Name      & RA          & Dec.    & z   & log(M$_{BH}/M_{\odot}$) & log(L$_{5100}$)          \\
{(A)}    & {(B)}        & {(C)}  &    {(D)}     &   {(E)}     &   {(F)}          \\
\hline

Mrk  335         & 00 06 19.44   & +20 12 07.2   & 0.026  & 7.23 & 43.76   \\
PG 0026+129      & 00 29 13.44    & +13 16 01.2  & 0.142    & 8.48 & 44.97  \\
PG 0052+251      & 00 54 52.08  & +25 25 37.2 & 0.154  & 8.46 & 44.81   \\
NGC 4253         & 12 18 26.40  & +29 48 43.2  &0.013  & 6.82 & 42.57   \\
PG 1307+085      & 13 09 46.80 & +08 19 48.0    & 0.155    & 8.53 & 44.85   \\

RMID733$^{*}$         & 14 07 59.04 & +53 47 56.4  & 0.455    & 8.20        & 43.40    \\
RMID399$^{*}$         & 14 10 31.20 & +52 15 32.4 & 0.608    & 8.10         & 44.10    \\
RMID101$^{*}$          & 14 12 14.16 & +53 25 44.4  & 0.458    & 7.90         & 43.40   \\
PG 1411+442      & 14 13 48.24   & +44 00 10.8 & 0.089   & 8.54 & 44.56   \\
RMID779$^{*}$          & 14 19 23.28  & +54 22 01.2  & 0.152    & 7.40         & 42.60    \\
RMID300$^{*}$          & 14 19 41.04 & +53 36 46.8  & 0.646    & 8.20         & 44.00      \\
Mrk  817         & 14 36 22.08  & +58 47 38.4  & 0.031  & 7.58 & 43.74   \\
Mrk  290         & 15 35 52.08 & +57 54 07.2  & 0.029  & 7.27 & 43.17   \\
Mrk  486         & 15 36 38.16  & +54 33 32.4  & 0.039 & 7.63 & 43.69   \\
Mrk  493          & 15 59 09.36 & +35 01 44.4  & 0.031 & 6.19  & 43.11   \\
PG 1613+658      & 16 13 57.12  & +65 43 08.4  & 0.129    & 8.33 & 44.77   \\
PG 1617+175     & 16 20 11.28 & +17 24 25.2 & 0.112 & 8.66 & 44.39   \\
1RXS J1858+4850 & 18 58 00.96 & +48 50 20.4  & 0.079    & 6.70  & 43.65   \\
PG 2130+099     & 21 32 27.60  & +10 08 16.8  & 0.063  & 7.43  & 44.20  \\

\hline
\end{tabular}

\vspace{0.2cm}
{\small {\bf Note}: Column (A) denotes the common name of the sources,  Column (B) and (C) present the right ascension (RA) and declination (Dec.) of the sources as obtained from the literature. Column (D) is the redshift value. Column (E) denotes the SMBH mass in the units of $M_{\odot}$, and  Column (F) is the luminosity at 5100\AA{}. The information for all the parameters is obtained from the AGN Black Hole Mass database \citep{Bentz2015} except for the sources marked with ${*}$ which are taken from \citet{Grier2018}.}
\end{table}

\begin{figure}
 
    \centering
    \includegraphics[width=9cm,height=8cm]{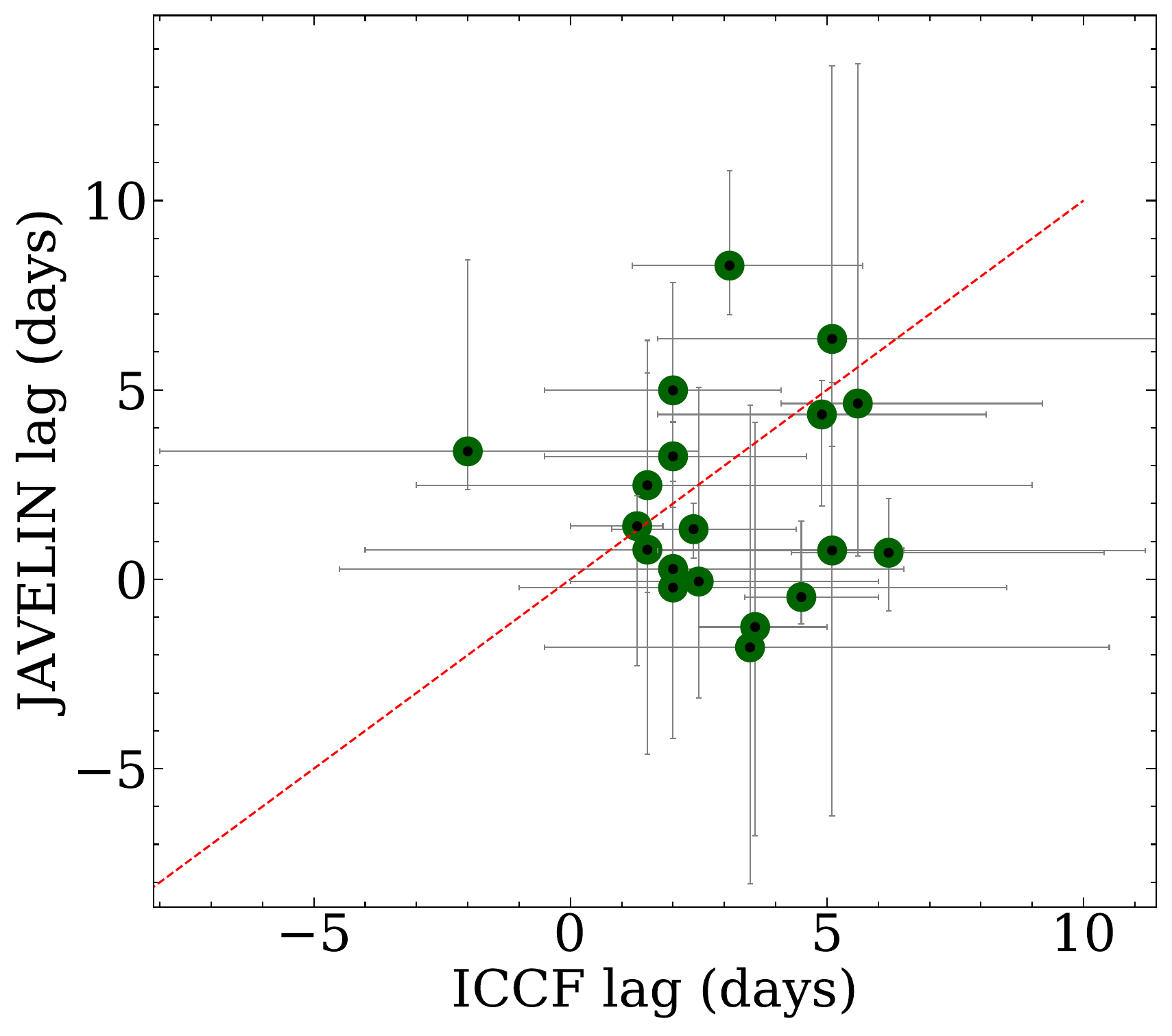}
    \caption{Comparison of the $g-i$ band lags obtained using the {\sc javelin}  and ICCF methods for the 19 sources used in this study. The red dashed line indicates a one-to-one correlation between the lags obtained using both the methods. }
 \label{Figure3}
\end{figure}

\begin{figure*}
    
    \subfigure{\includegraphics[width=6cm,height=5cm]{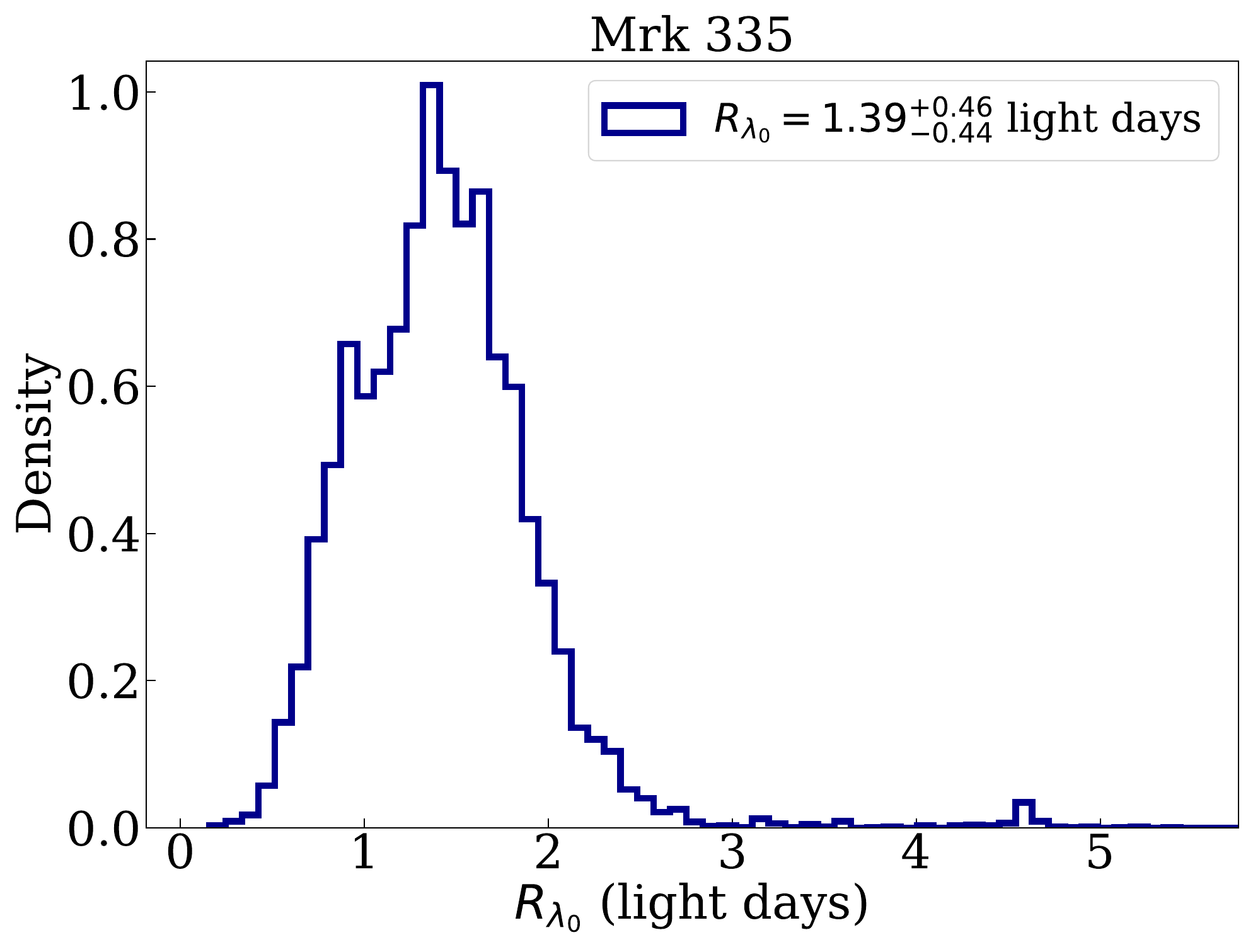}}
    \subfigure{\includegraphics[width=6cm,height=5cm]{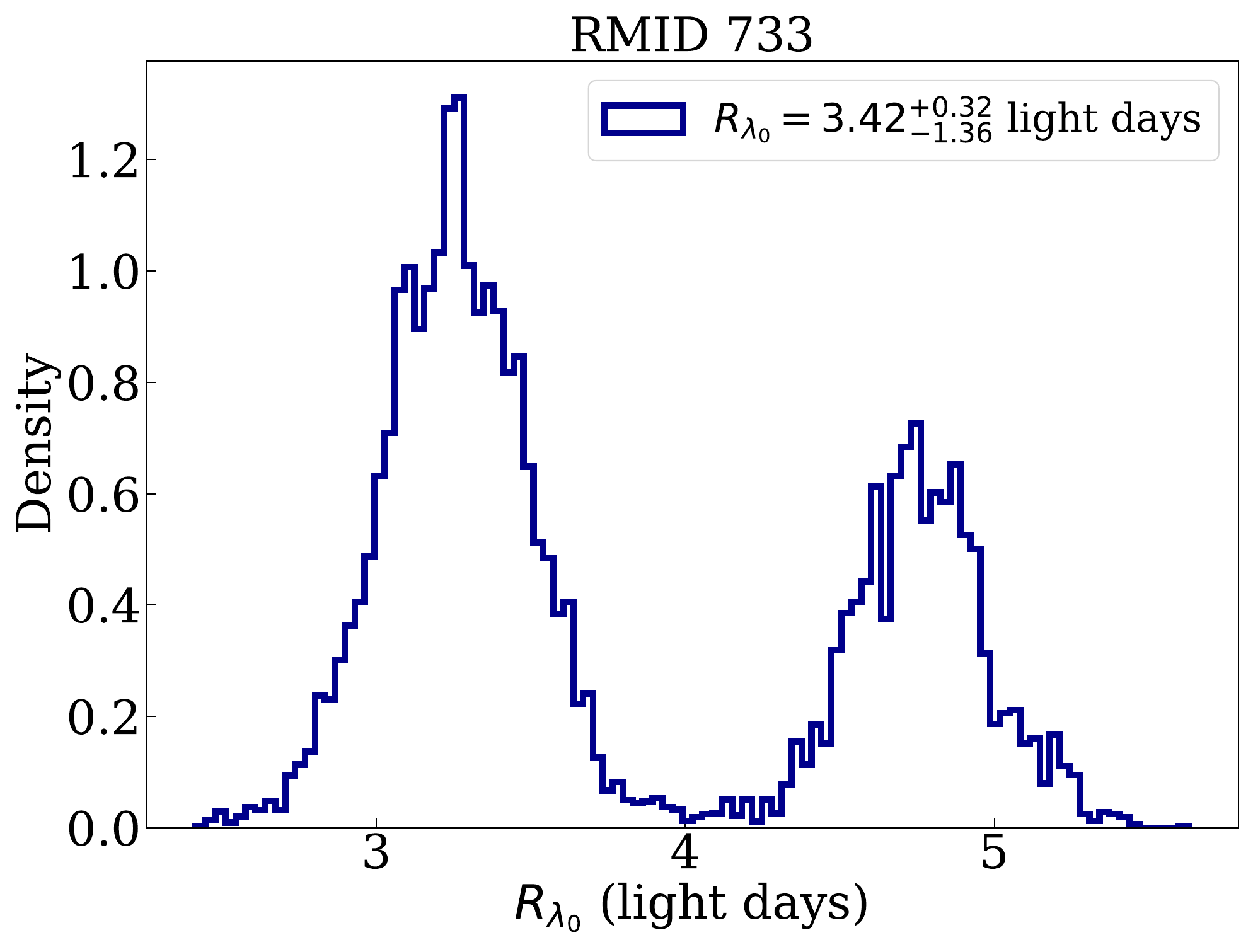}} \\
    \subfigure{\includegraphics[width=6cm,height=5cm]{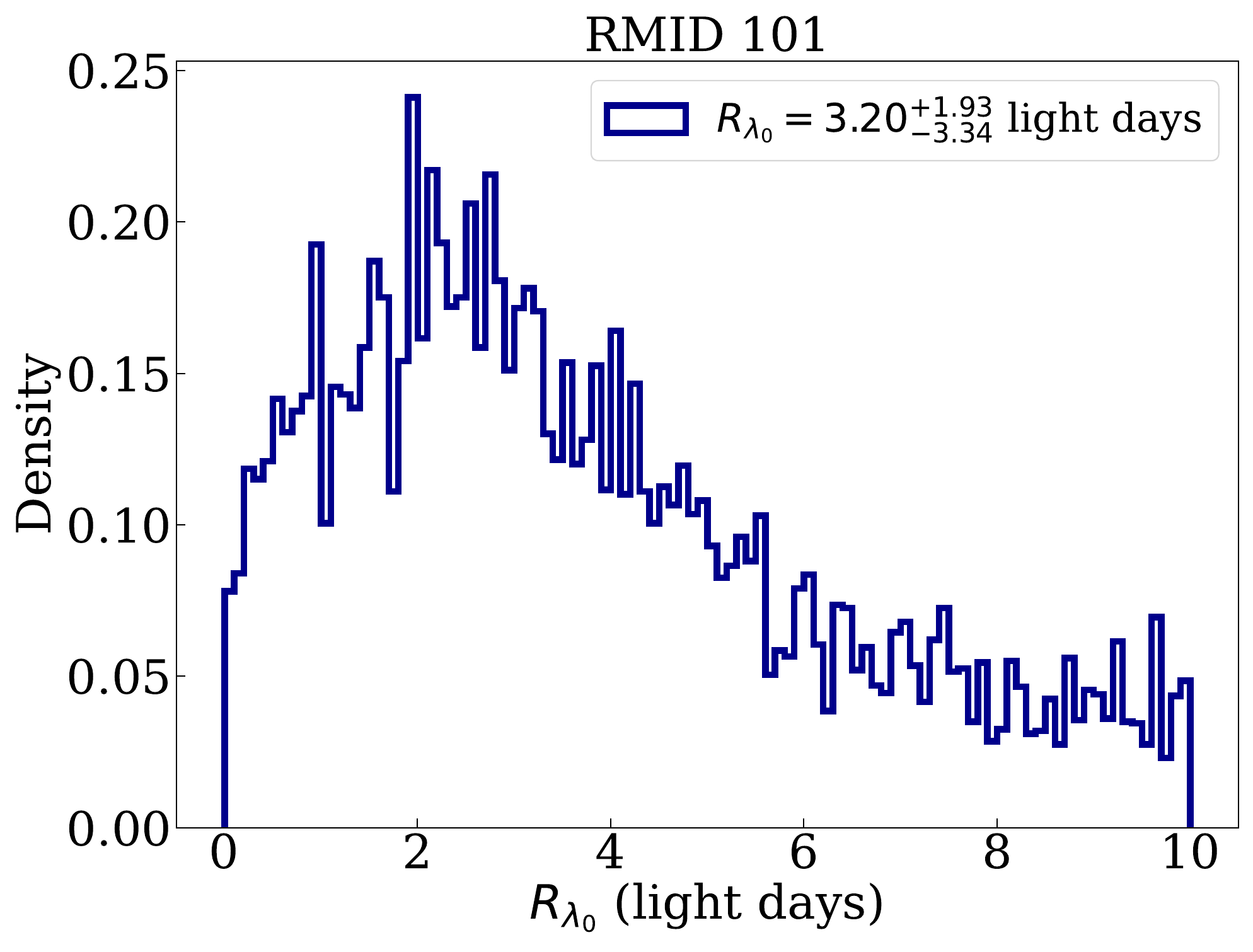}}
    \subfigure{\includegraphics[width=6cm,height=5cm]{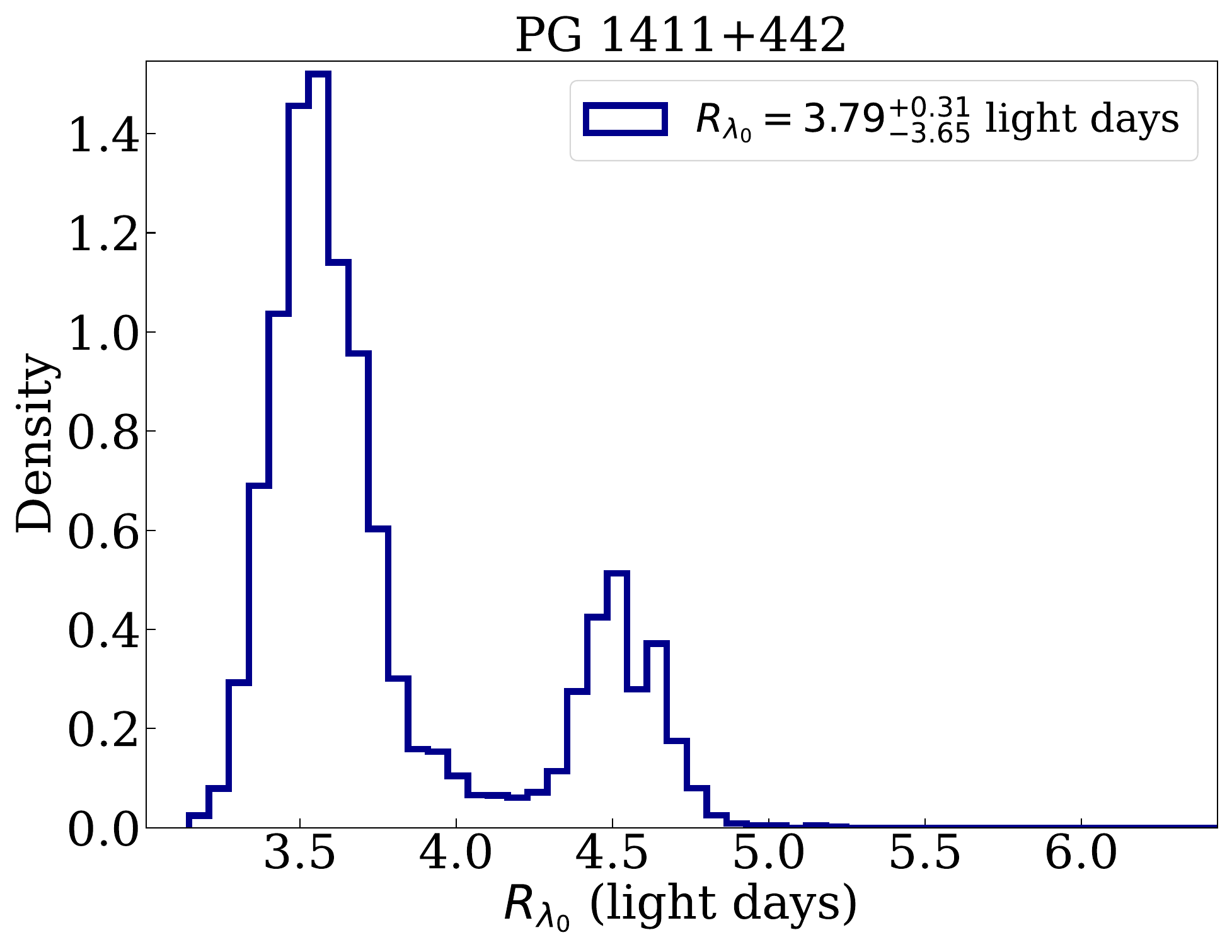}}  \\
    \subfigure{\includegraphics[width=6cm,height=5cm]{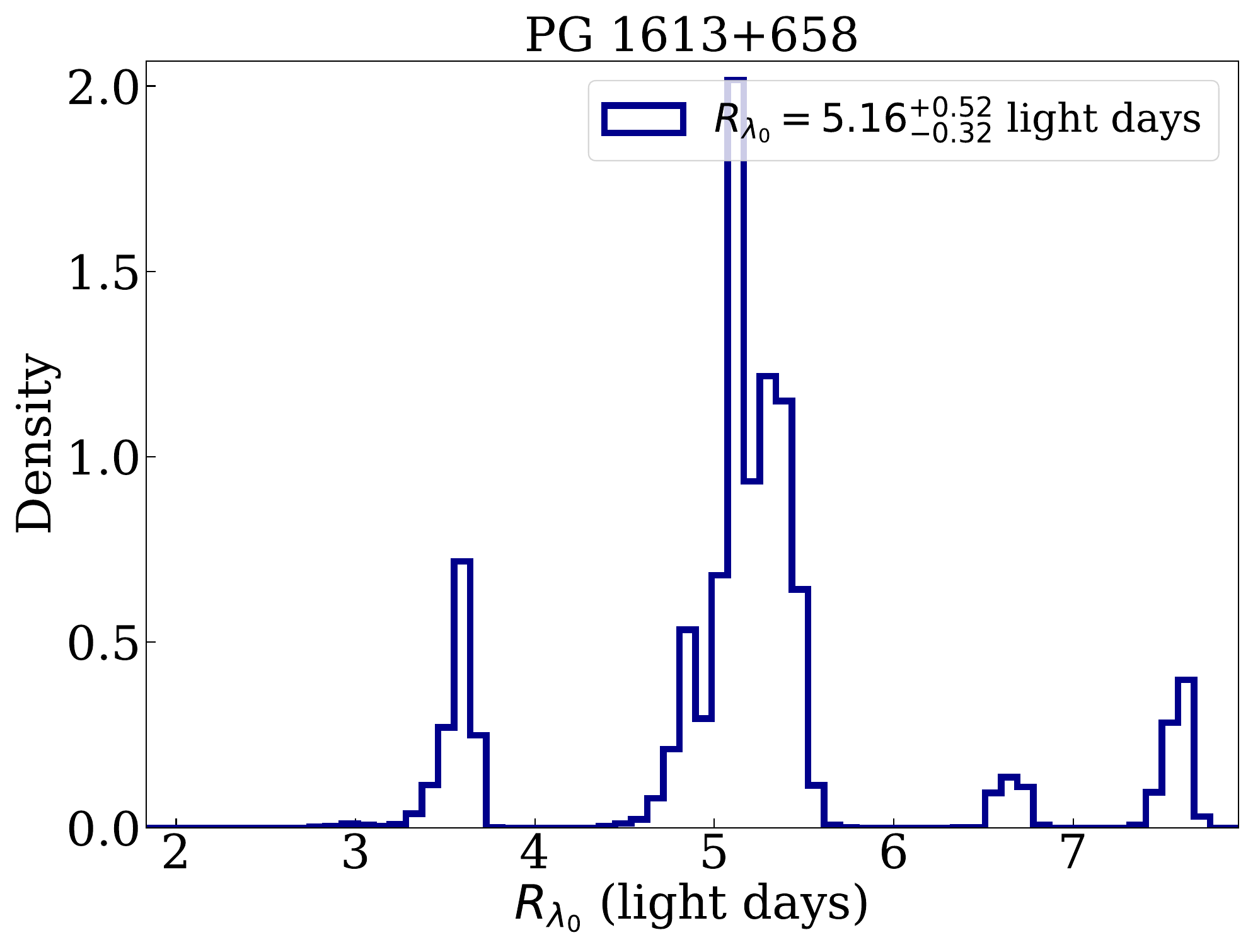}}
    \subfigure{\includegraphics[width=6cm,height=5cm]{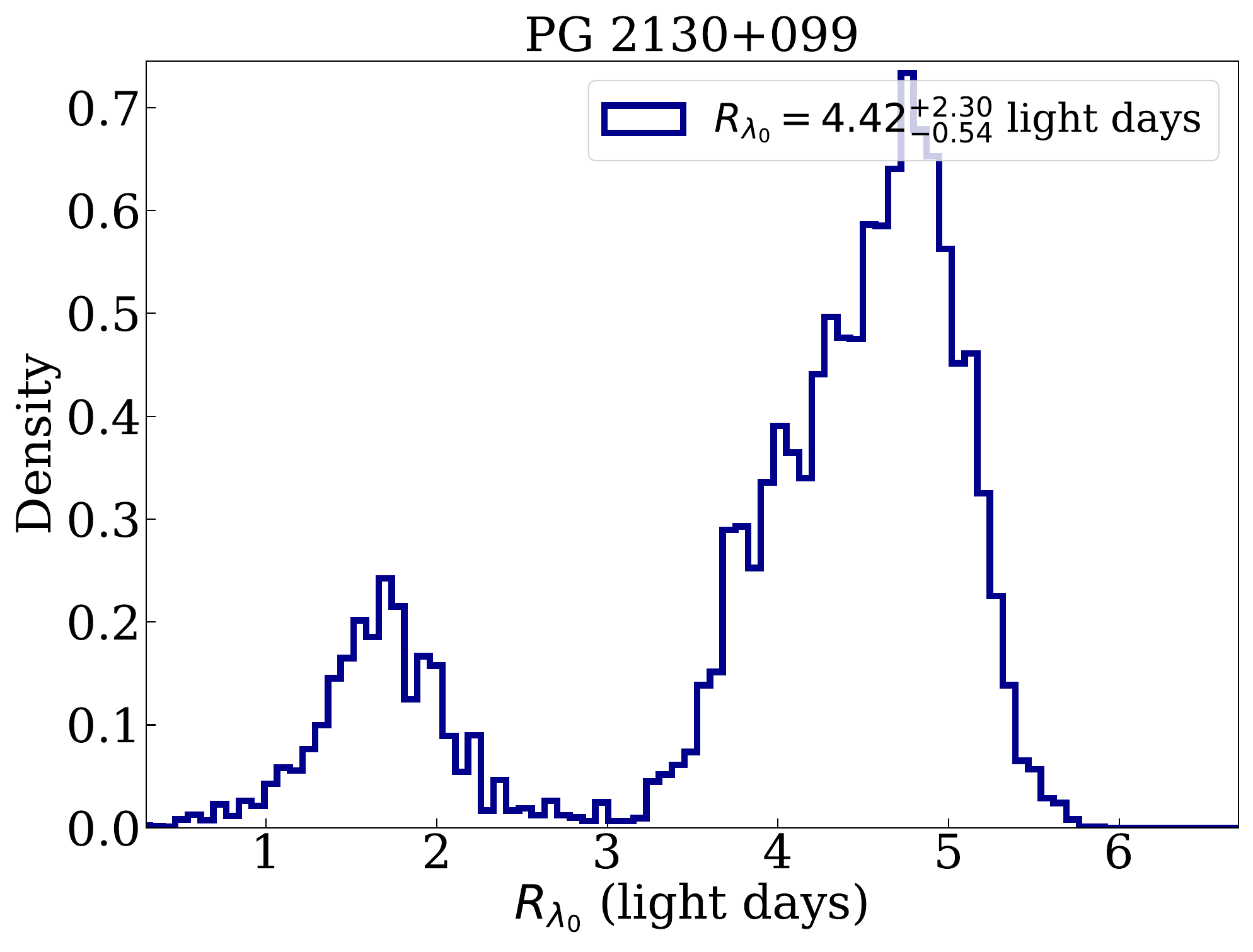}}

    \caption{Distribution for the disk size at  $g$-band rest wavelength ($R_{\lambda_0}$) for 6 sources from our sample are shown here. This Distribution for $R_{\lambda_0}$ is obtained using the {\sc javelin}  thin disk model assuming disk size scaling with temperature as a power-law index of 4/3. For most of the sources, the estimated $R_{\lambda_0}$ is larger than the predictions of the SS disk model (see columns F and G of Table \ref{Table3}).}
 \label{Figure5}  
\end{figure*}

\begin{table}

\fontsize{6}{8.0}\selectfont
\caption{Light curve statistics for the sample of 19 AGN with reasonable lag estimates.}
\label{Table2}
\begin{tabular}{lrrrrrr}
\hline
\hline   
& \multicolumn{2}{c}{ $g$-band}                              & \multicolumn{2}{c}{r-band}   & \multicolumn{2}{c}{i-band}                              \\
 \hline
Name            & nobs & Mag & nobs & Mag & nobs & Mag \\
{(A)}  & {(B)}  & {(C)} &  {(D)} &  {(E)} &   {(F)} & {(G)}          \\
\hline

Mrk  335 (S1)         & 48     & 14.82        $\pm 0.01$       & 52     & 14.28         $\pm 0.01$      & 27     & 14.34         $\pm 0.01$       \\
Mrk  335 (S2)         & 72     & 14.80        $\pm 0.01$       & 67     & 14.25         $\pm 0.01$      & 24     & 14.27         $\pm 0.01$       \\
PG0026+129 (S2)      & 70     & 15.31        $\pm 0.01$       & 120    & 15.11         $\pm 0.01$      & 19     & 14.68         $\pm 0.01$      \\  
PG0052+251 (S1)      & 43     & 16.21        $\pm 0.01$       & 52     & 16.02         $\pm 0.01$      & 24     & 15.36         $\pm 0.01$       \\
PG0052+251 (S2)      & 68     & 16.09        $\pm 0.01$       & 67     & 15.93         $\pm 0.01$      & 18     & 15.31         $\pm 0.01$       \\
NGC4253 (S1)         & 304    & 15.07        $\pm 0.01$       & 337    & 14.46         $\pm 0.01$      & 45     & 14.32         $\pm 0.02$       \\
NGC4253 (S2)         & 168    & 15.17        $\pm 0.01$       & 143    & 14.51         $\pm 0.01$      & 29     & 14.35         $\pm 0.02$       \\
PG1307+085 (S1)      & 49     & 16.00        $\pm 0.01$       & 102    & 15.83         $\pm 0.01$      & 20     & 15.35         $\pm 0.01$       \\
PG1307+085 (S2)      & 63     & 15.70        $\pm 0.01$       & 72     & 15.59         $\pm 0.01$      & 19     & 15.21         $\pm 0.01$       \\
RMID733 (S1)         & 231    & 18.34        $\pm 0.04$       & 331    & 17.85         $\pm 0.03$      & 25     & 17.46         $\pm 0.03$       \\
RMID399 (S1)         & 113    & 20.77        $\pm 0.19$       & 237    & 20.37         $\pm 0.15$      & 19     & 19.76         $\pm 0.13$       \\
RMID101 (S1)         & 227    & 18.84        $\pm 0.06$       & 325    & 18.81         $\pm 0.06$      & 23     & 18.72         $\pm 0.06$       \\
PG1411+442 (S1)      & 452    & 14.98        $\pm 0.01$       & 447    & 14.75         $\pm 0.01$      & 41     & 14.66         $\pm 0.01$       \\
PG1411+442 (S2)      & 196    & 14.97        $\pm 0.02$       & 211    & 14.75         $\pm 0.01$      & 32     & 14.65         $\pm 0.02$       \\
RMID779 (S1)         & 340    & 19.69        $\pm 0.14$       & 321    & 19.26         $\pm 0.09$      & 24     & 18.83         $\pm 0.07$       \\
RMID300 (S1)         & 392    & 19.24        $\pm 0.09$       & 318    & 19.20         $\pm 0.07$      & 21     & 19.23         $\pm 0.09$       \\
Mrk  817 (S1)         & 227    & 14.31        $\pm 0.01$       & 246    & 14.08         $\pm 0.01$      & 25     & 14.15         $\pm 0.01$       \\
Mrk  817 (S2)         & 93     & 14.36        $\pm 0.01$       & 102    & 14.11         $\pm 0.01$      & 16     & 14.16         $\pm 0.01$       \\
Mrk  290 (S1)         & 375    & 15.10        $\pm 0.01$       & 412    & 14.72         $\pm 0.01$      & 28     & 14.78         $\pm 0.01$       \\
Mrk  486 (S1)         & 377    & 14.78        $\pm 0.01$       & 432    & 14.28         $\pm 0.01$      & 27     & 14.32         $\pm 0.01$       \\
Mrk  486 (S2)         & 144    & 14.80        $\pm 0.01$       & 188    & 14.28         $\pm 0.01$      & 21     & 14.31         $\pm 0.01$       \\
Mrk 493 (S1)          & 681    & 15.77        $\pm 0.01$       & 623    & 15.33         $\pm 0.01$      & 47     & 15.22         $\pm 0.01$       \\
Mrk 493 (S2)          & 297    & 15.77        $\pm 0.01$       & 236    & 15.31         $\pm 0.01$      & 49     & 15.18         $\pm 0.01$       \\
PG1613+658 (S1)      & 644    & 15.08        $\pm 0.02$       & 615    & 15.07         $\pm 0.01$      & 48     & 14.51         $\pm 0.01$       \\
PG1613+658 (S2)      & 335    & 15.45        $\pm 0.02$       & 318    & 15.38         $\pm 0.01$      & 53     & 14.74         $\pm 0.01$       \\
PG 1617+175 (S2)     & 40     & 15.49        $\pm 0.01$       & 78     & 15.49         $\pm 0.01$      & 25     & 15.01         $\pm 0.01$       \\
1RXS J1858+4850 (S1) & 452    & 16.57        $\pm 0.01$       & 728    & 16.22         $\pm 0.01$      & 28     & 16.27         $\pm 0.01$       \\
1RXSJ1858+4850 (S2)  & 288    & 16.66        $\pm 0.01$       & 276    & 16.32         $\pm 0.01$      & 28     & 16.36         $\pm 0.01$       \\
PG 2130+099 (S1)     & 63     & 14.54        $\pm 0.01$       & 68     & 14.27         $\pm 0.01$      & 26     & 14.38         $\pm 0.01$       \\
PG2130+099 (S2)      & 41     & 14.65        $\pm 0.01$       & 37     & 14.37         $\pm 0.01$      & 25     & 14.51         $\pm 0.01$       \\

\hline

\end{tabular}

\vspace{0.2cm}
{\small {\bf Note}: Column (A) denotes the common name of the sources as available in  \citet{Bentz2015} and \citet{Grier2018}. S1 and S2 denote the light curves for the two seasons used in this study. Columns (B), (D), and (F) are the number of observations (nobs) in the $g$, $r$, and $i$ filters, respectively, as available in ZTF-DR6. The mean magnitudes and their uncertainties in the $g$, $r$, and $i$  filters (Mag) throughout the observations are presented in columns (C), (E), and (G).}


\end{table}

\begin{table*}
\caption{ The lags and the disk size estimates along with the SS disk prediction for 19 sources in our sample.}
\label{Table3}
\setlength{\tabcolsep}{9pt} 
\renewcommand{\arraystretch}{1.5} 

\begin{tabular}{lrrrrrrr}
\hline
Common name  & $\tau_r$ ({\sc javelin} ) &  $\tau_i$ ({\sc javelin} ) & $\tau_r$  (ICCF) & $\tau_i$ (ICCF) & $R_{0}$ ( $g$-band) & $R_{0}$ (SS disk) &\\
    & (days)      & (days)   &    (days)     &  (days)      &   (light-days)      &   (light-days)     &     \\

 {(A)}    & {(B)}        & {(C)}  &    {(D)}     &   {(E)}     &   {(F)}     &   {(G)}     &     \\

\hline

Mrk  335 (S1)       & $0.94^{+0.45}_{-1.60}$ & $1.32^{+0.76}_{-0.68}$  & $1.5^{+1.2}_{-1.3}$ & $2.4^{+1.6}_{-2.0}$ & $1.39^{+0.46}_{-0.44}$ & 1.33 &  \\
PG 0026+129 (S2)    & $3.30^{+3.70}_{-7.62}$   & $2.48^{+2.82}_{-2.95}$ &  $0.6^{+0.9}_{-0.2}$ & $2.0^{+2.5}_{-2.6}$ & $6.29^{+1.24}_{-0.87}$& 2.81 &\\
PG 0052+251 (S2)   & $2.88^{+1.75}_{-2.37}$  & $-1.79^{+6.24}_{-6.40}$ & $2.5^{+2.5}_{-0.0}$ & $3.5^{+4.0}_{-7.0}$ & $2.79^{+1.98}_{-3.26}$ & 2.44 &  \\
NGC 4253 (S1)     & $-0.01^{+2.06}_{-1.98}$  & $-0.22^{+3.98}_{-2.79}$ & $0.5^{+3.9}_{-2.0}$ & $-2.0^{+3.0}_{-6.5}$  &$ 5.98^{+5.89}_{-2.09}$ & 0.55 & \\
PG 1307+085 (S2)        & $0.10^{+7.27}_{-0.46}$  & $-0.05^{+3.07}_{-5.12}$ &$3.0^{+2.6}_{-2.6}$ & $1.5^{+4.5}_{-7.5}$& $9.16^{+4.96}_{-0.62}$ & 2.52& \\
RMID733 (S1)      & $0.92^{+0.47}_{-2.47}$  & $3.37^{+1.00}_{-5.05}$ & $-0.5^{+1.9}_{-2.5}$& $-2.0^{+6.0}_{-4.5}$  & $3.42^{+0.32}_{-1.36} $ & 0.56 & \\
RMID399 (S1)        & $2.62^{+5.12}_{-5.59}$  & $4.64^{+4.03}_{-8.97}$ & $6.8^{+2.2}_{-10.3}$ & $5.6^{+1.5}_{-3.6}$ & $5.91^{+3.51}_{-3.11}$ & 0.80 &   \\
RMID101 (S1)  & $-2.95^{+5.13}_{-5.28}$  & $0.78^{+5.39}_{-5.52}$ & $3.3^{+4.1}_{-6.2}$ & $1.5^{+5.5}_{-5.0}$ & $3.20^{+1.93}_{-3.34}$ & 0.56 &  \\
PG 1411+442 (S2)   & $-3.67^{+0.67}_{-0.82}$  & $-0.46^{+0.70}_{-2.00}$  & $1.5^{+7.0}_{-6.4}$ & $2.5^{+2.5}_{-3.5}$ & $3.79^{+0.31}_{-3.65}$ & 2.22 & \\
RMID779 (S1)         & $4.56^{+3.90}_{-3.04}$  & $6.34^{+2.83}_{-7.21}$  & $9.0^{+0.0}_{-2.8}$ &$5.1^{+3.4}_{-9.1}$ & $6.21^{+2.91}_{-2.82}$ & 0.46 &  \\
RMID300 (S1) &  $3.77^{+1.62}_{-6.41}$ & $0.75^{+7.00}_{-4.43}$ & $2.0^{+2.6}_{-3.0}$ &  $2.0^{+6.5}_{-4.5}$ &  $7.84^{+3.39}_{-1.56}$ & 0.71\\
Mrk  817 (S1)    &$ 4.31^{+0.21}_{-1.99}$  & $4.98^{+0.82}_{-2.89}$ & $2.5^{+1.5}_{-1.5}$ & $3.6^{+1.1}_{-1.4}$ & $4.70^{+0.18}_{-3.83}$ & 1.30 &  \\
Mrk  290 (S1)          &$0.84^{+3.28}_{-1.64}$  & $0.27^{+2.06}_{-1.62}$ &$4.1^{+1.9}_{-5.9}$ & $2.0^{+2.5}_{-2.1}$ & $5.14^{+2.05}_{-2.62}$ & 0.84 & \\
Mrk  486 (S2)         & $0.38^{+0.21}_{-0.36}$  & $4.35^{+2.41}_{-0.89}$ & $0.5^{+1.0}_{-1.0}$ & $4.5^{+1.1}_{-1.5}$  & $0.93^{+0.11}_{-2.05}$ &1.23 & \\
Mrk  493 (S2)         &$ 0.68^{+1.68}_{-0.29} $ & $0.70^{+1.53}_{-1.43}$   & $2.0^{+0.6}_{-0.5}$ & $4.9^{+3.2}_{-3.2}$ & $5.52^{+1.10}_{-0.98}$   & 0.81 &\\
PG 1613+658 (S1)        & $-0.38^{+5.86}_{-0.46}$  & $-1.26^{+5.51}_{-5.39}$ &  $2.0^{+1.9}_{-2.6}$ & $5.1^{+3.4}_{-6.1}$ &
$5.16^{+0.52}_{-0.32}$  & 2.46 &\\
PG 1617+175 (S2)       & $3.41^{+2.04}_{-1.98}$   & $8.28^{+1.29}_{-2.51}$ & $5.2^{+2.9}_{-1.4}$ & $6.2^{+1.9}_{-4.2}$ &
 $8.14^{+1.75}_{-1.31}$& 1.89 & \\
1RXS J1858+4850 (S2)         & $0.37^{+0.13}_{-0.91}$  & $1.40^{+3.68}_{-0.81}$ &
$1.0^{+1.1}_{-1.5}$ & $3.1^{+1.9}_{-2.6}$ &
$2.74^{+1.90}_{-0.25}$ &1.13 & \\
PG 2130+099 (S2)       & $2.67^{+0.73}_{-3.05}$  & $3.24^{+3.06}_{-0.90}$ & $1.2^{+0.6}_{-0.5}$ & $1.3^{+1.3}_{-0.5}$ & 
$4.42^{+2.30}_{-0.54}$  & 1.76 &\\

\hline
\end{tabular}

\vspace{0.2cm}
{{\bf Note}: Column (A) denotes the common name of the sources as available in \citet{Bentz2015} and \citet{Grier2018}. S1 and S2 denote the respective season in which the interband lags increasing with wavelength were obtained using either of the methods. The $g-r$ and $g-i$ reverberation lags obtained using {\sc javelin} are shown in Columns (B) and (C), while the lags obtained using ICCF method for $g-r$ and $g-i$ are presented in columns (D) and (E) respectively. The disk sizes obtained using the {\sc javelin} thin disk model are presented in column (F) and the expected theoretical disk sizes at $g-$band effective wavelength assuming the SS disk model are presented in column (G).}

\end{table*}

\section{Accretion disk sizes}
\label{section5}
\subsection{Disk sizes using {\sc javelin}  thin disk model}

A recent extension in the {\sc javelin}  code known as the {\it thin disk model} introduced by \citet{Mudd2018} allows us to directly model and estimate the size of the accretion disk at a particular wavelength based on the multi-band light curves. It fits the light curves assuming the SS disk approximation, and the accretion disk size is given according to equation \ref{equation2}. 

This method was used by \citet{Mudd2018} to obtain the accretion disk size at 2500\AA{} for a set of 15 quasars from the DES survey. Further, \citet{Kokubo2018} and \citet{Yu2018} have also used this model to derive the accretion disk sizes for different quasars. We used the thin disk model for the light curves to obtain the disk sizes at  $g$-band rest wavelength and compare it to the predicted SS disk model based on individual SMBH mass and accretion rates. It is noticeable that the disk sizes predicted for {\sc javelin}  assume the SS disk as we have fixed $\beta$ to 1.33 (4/3). Leaving the parameter $\beta$ free does not converge to any particular value, as noticed in the recent works \citep[][]{Kokubo2018, Yu2018}. However, a possible concern is that this model could be unable to recover the accurate parameters for AGN accreting at a faster rate which may require additional physical processes or special geometry to be explained \citep{castello2017}.

The distribution for $R_{\lambda_0}$ gives the size of the accretion disk at the respective wavelength $\lambda_0$. Since the sources in our sample were located at varying redshifts, the corresponding wavelength ($\lambda_0$) for the disk size varied from 2717 \AA{} to 4677 \AA{} according to the  $g$-band rest-frame wavelength. The values for $R_{\lambda_0}$ ranged from 0.93 light days to 9.16 light days with a mean value of 4.88 light days. Since the sources for this analysis have large SMBH mass and luminosity ranges, it is plausible that the estimated disk sizes vary from source to source. The distribution for $R_{\lambda_0}$ for a few sources is shown in Figure \ref{Figure5}. 

\subsection{ Comparison with analytical SS disk model}

The standard SS disk model predicts the disk sizes based on the SMBH mass, and the mass accretion rate. We compare the disk sizes obtained using this analytical model with the disk sizes obtained by fitting the light curves through the {\sc javelin}  thin disk model. Following \citet{Jiang2017}, the
light travel time across two different radii where photons with
wavelengths $\lambda_g$ and $\lambda_x$ are emitted, can be put as:

\begin{equation}
\centering
\Delta t_{g-x}=\left(X\frac{k_{\rm B}\lambda_g}{hc}\right)^{4/3}\left(f_i\frac{3GM_{BH}\dot{M}}{8\pi\sigma}\right)^{1/3}
\left[\left(\frac{\lambda_x}{\lambda_g}\right)^{4/3}-1\right]
\end{equation}

Where, $k_B$, h, c, G and $\sigma$ are the Boltzmann constant, Planck constant, the speed of light, the gravitational constant and Stefan's constant respectively, $M_{BH}$ is the SMBH mass, $\dot{M}$ is the mass accretion rate, and X is a scaling factor. The value of X has been chosen to be 2.49 as per the recent works \citep[see][]{Fausnaugh2016, Edelson2019}.

\begin{figure*}

    \centering
    \includegraphics[width=8cm,height=7cm]{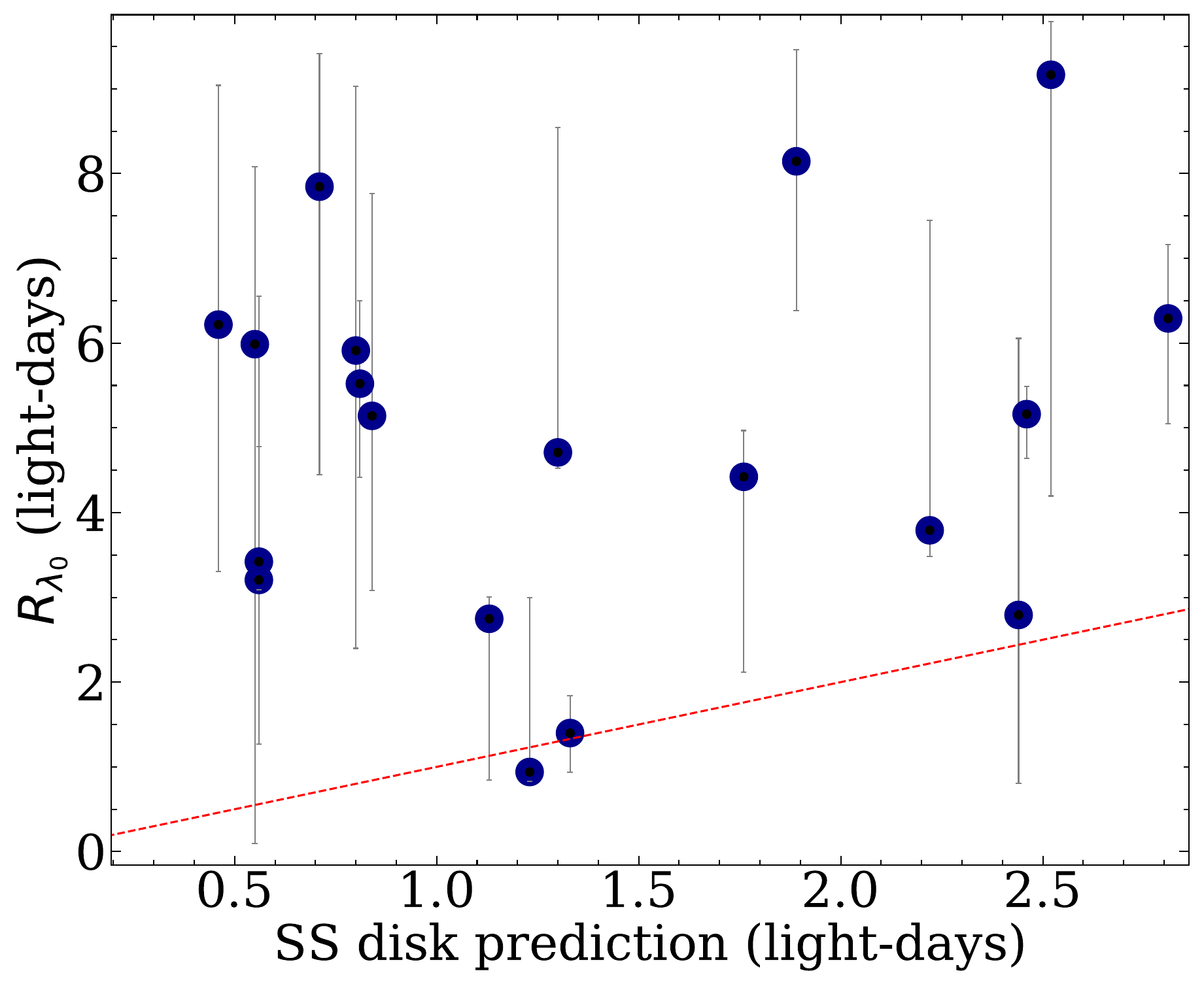}
    \caption{Comparison of the disk sizes ($R_{\lambda_0}$) at  $g$-band rest wavelength. The estimation for $R_{\lambda_0}$ has been made using the {\sc javelin}  thin disk model, while the analytical disk size has been calculated using the SS disk model dependent on the SMBH mass and accretion rates. The red dashed line indicates a one-to-one correlation between the two estimates.}
 \label{Figure4}   
\end{figure*}

The disk size estimated by modeling the light curves through the {\sc javelin}  thin disk model should agree with the lags if the SS disk assumption is held valid. Figure \ref{Figure4} shows the comparison of the disk sizes obtained using the {\sc javelin}  thin disk model and the prediction of the SS disk model. For 16 sources in our sample, the obtained disk sizes are larger than the theoretical disk size estimates, which has been observed in previous works as well \citep[see][and references therein]{Shapee2014, Fausnaugh2016, Starkey, Edelson2019}. However, for 3 sources, the disk sizes agree with the SS disk predictions. For instance, our estimated disk size of  $1.39^{+0.46}_{-0.44}$ days at 4400 \AA{} for Mrk  335 is in close agreement with the disk size of 1.33 light days predicted based on standard SS disk at the same wavelength.

\begin{figure*}

\subfigure[]{   \includegraphics[width=7cm,height=5cm]{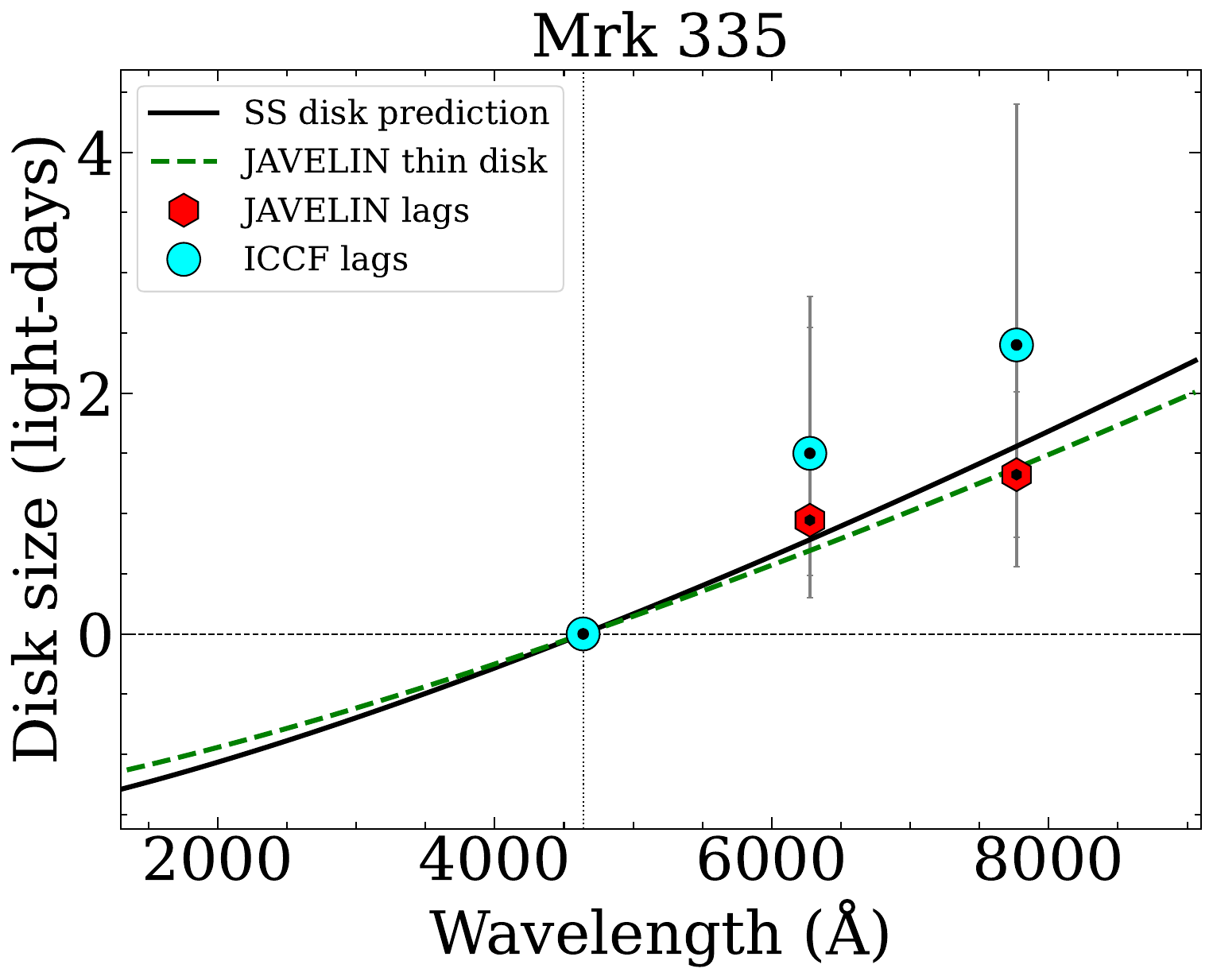}}
\subfigure[]{   \includegraphics[width=7cm,height=5cm]{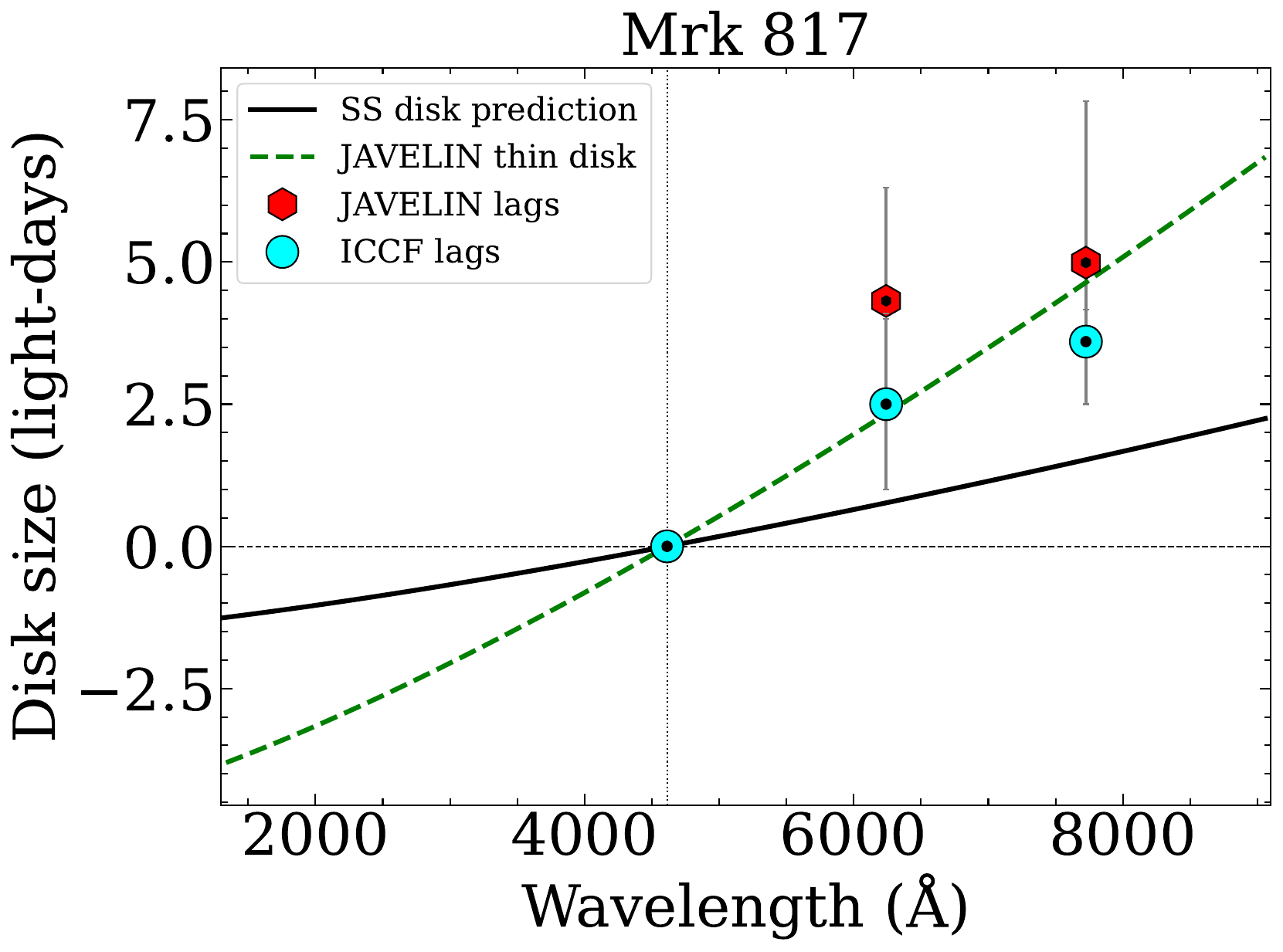}}
\subfigure[]{  \includegraphics[width=7cm,height=5cm]{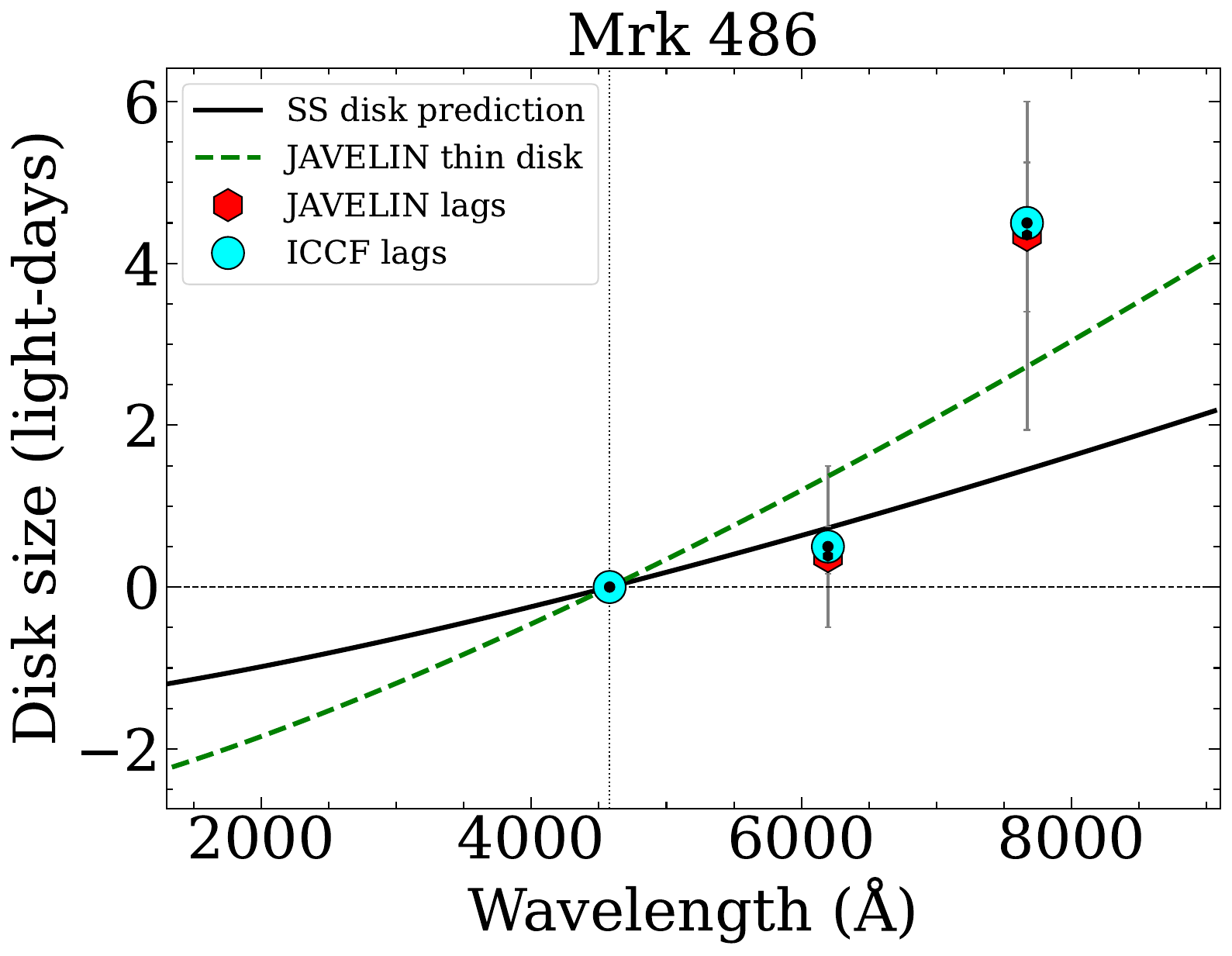}}
\subfigure[]{  \includegraphics[width=7cm,height=5cm]{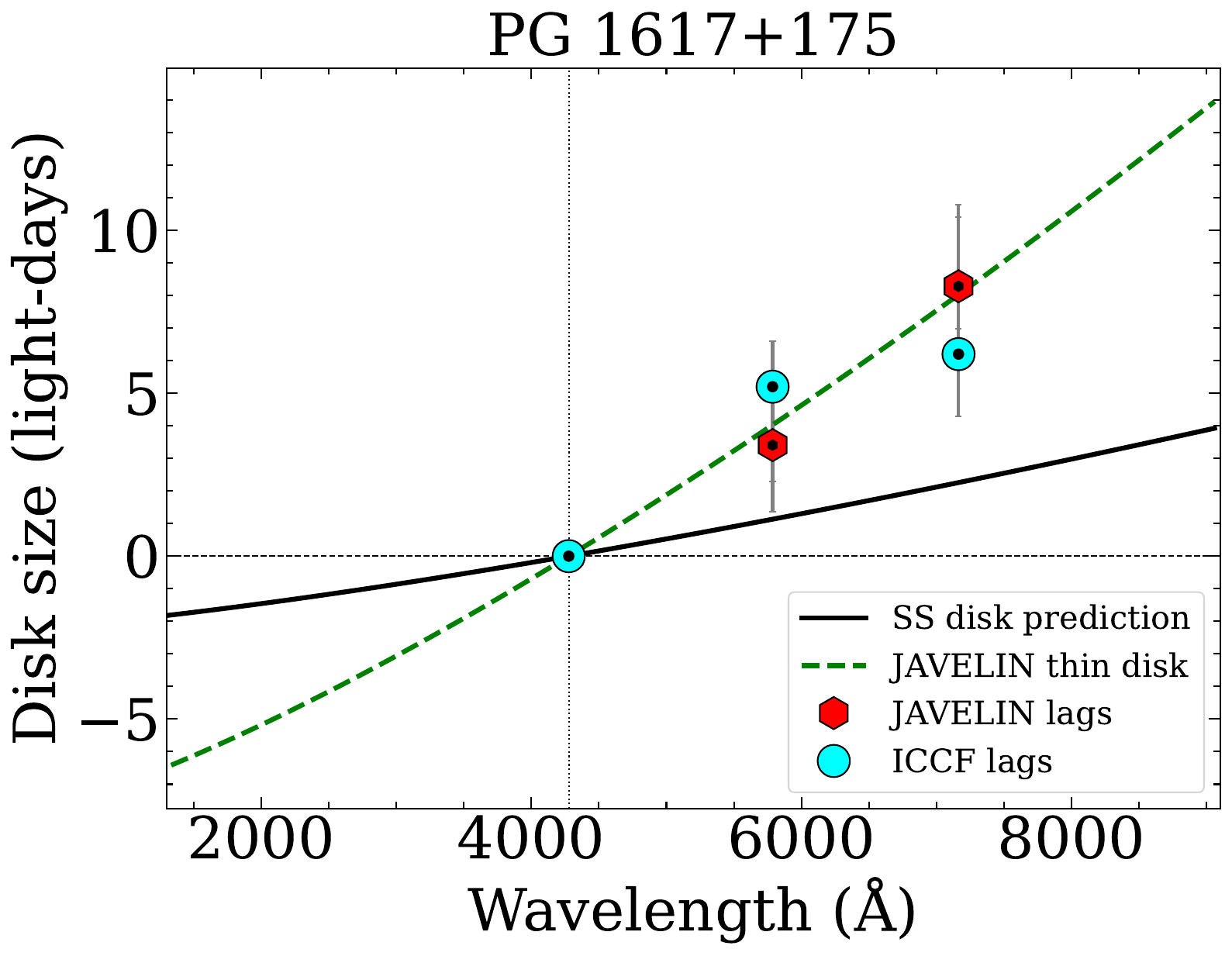}}
\subfigure[]{  \includegraphics[width=7cm,height=5cm]{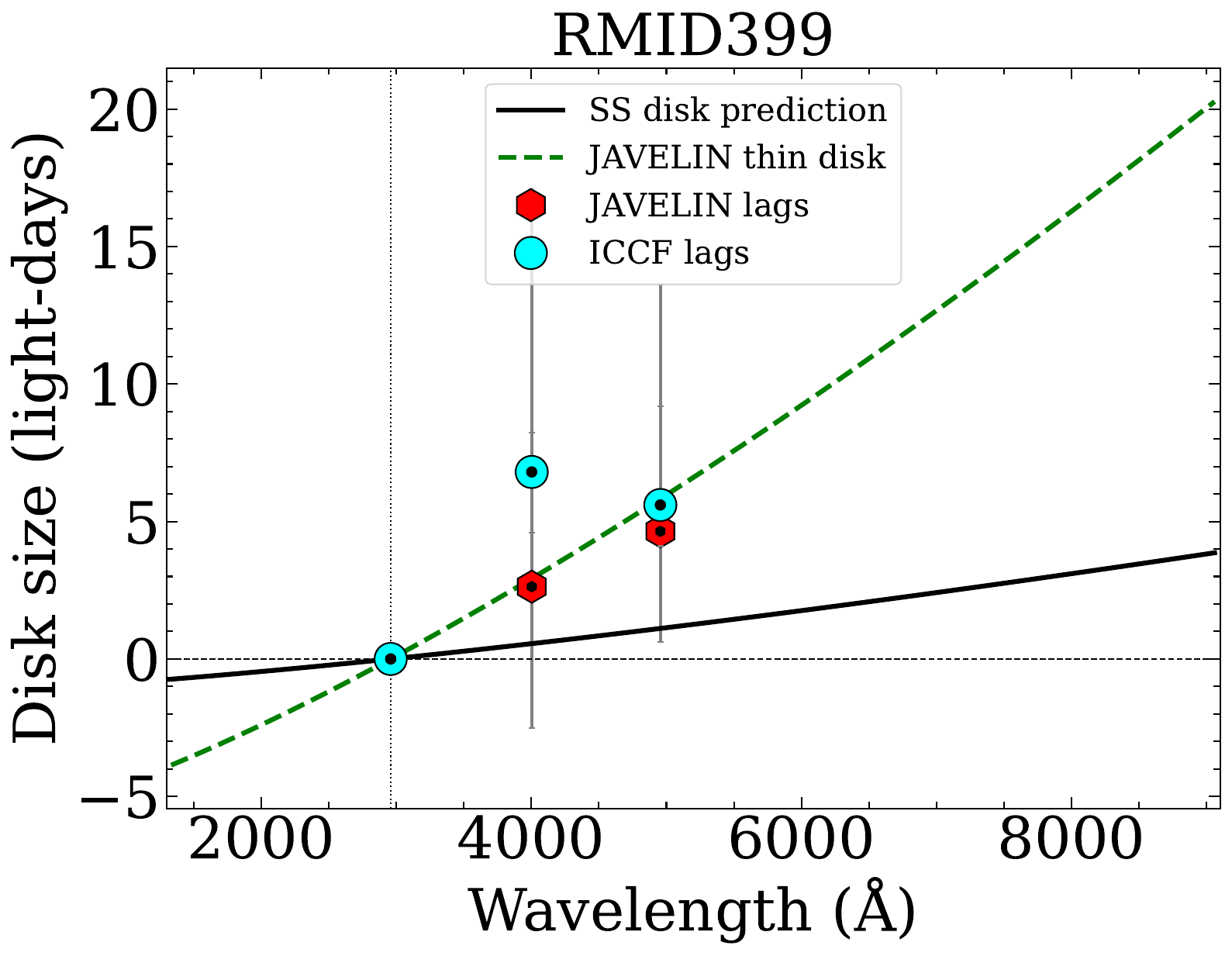}}
\subfigure[]{  \includegraphics[width=7cm,height=5cm]{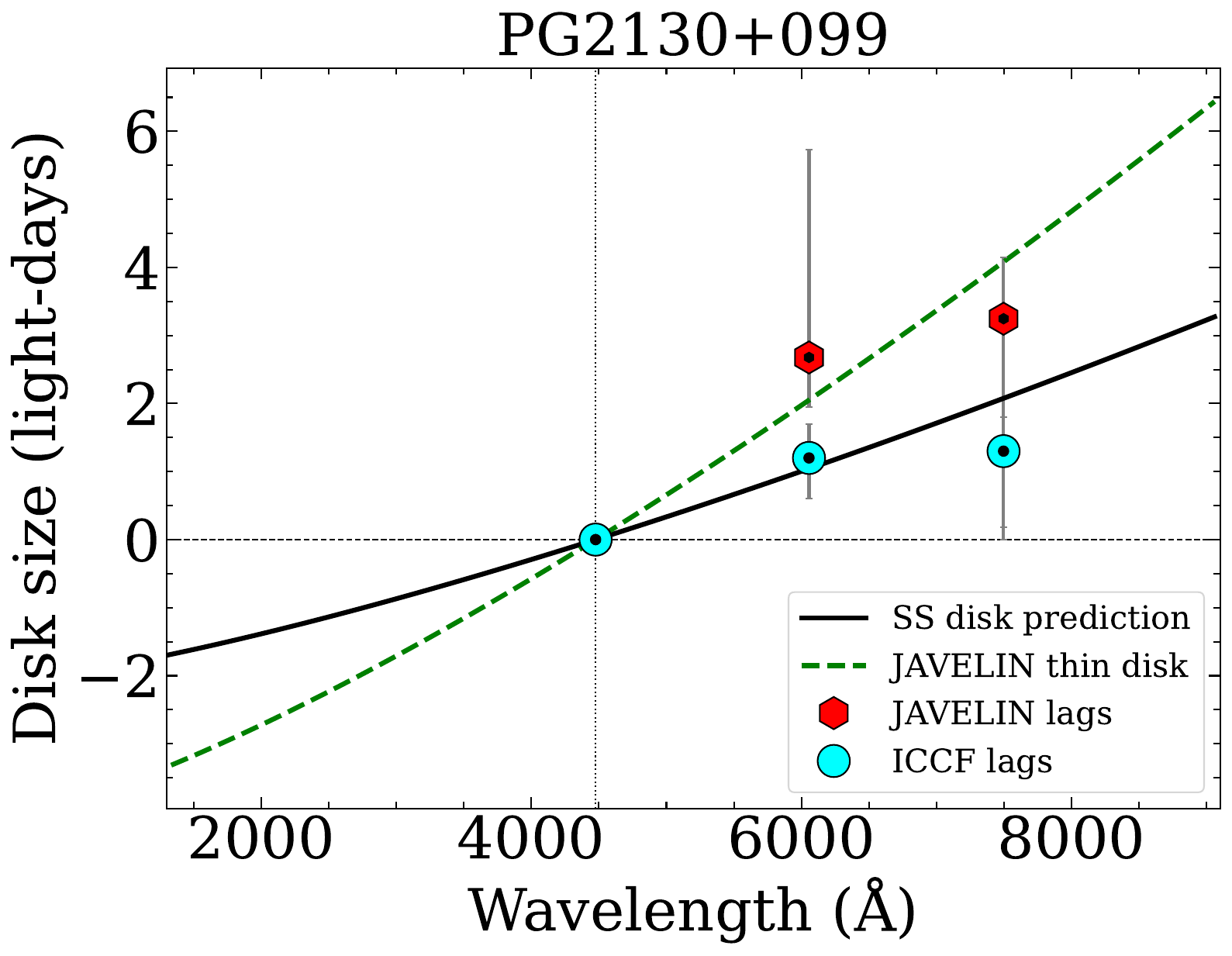}}

\caption{The SS disk prediction (black line) and the disk sizes as obtained through the {\sc javelin}  thin disk model (green dashed line) for 6 sources in our sample are shown here. Also, the $g-r$ and $g-i$ lags estimated using {\sc javelin}  and ICCF are shown in red and cyan colors, respectively. The results for all the sources with lag estimates constrained between [$-$10, 10] days are made available as supplementary material online.}

\label{Figure7}

\end{figure*}

We calculated the lag spectrum based on the predicted SS disk model, the obtained thin disk sizes by modeling the light curves using the {\sc javelin}  thin disk model, and the $g-r$ and $g-i$ lags obtained using the {\sc javelin}   and ICCF methods. We had only 2 interband lag measurements, while we used the combination of $R_{\lambda_0}$ obtained using the {\sc javelin}  thin disk model and the equation \ref{equation2} to compare the accretion disk sizes at different wavelengths. We find out that most sources have disk sizes and interband lags larger than the SS disk prediction. The results are shown in Figure \ref{Figure7}.

\begin{figure}
\hspace{-0.5cm}
\includegraphics[width=9cm,height=6cm]{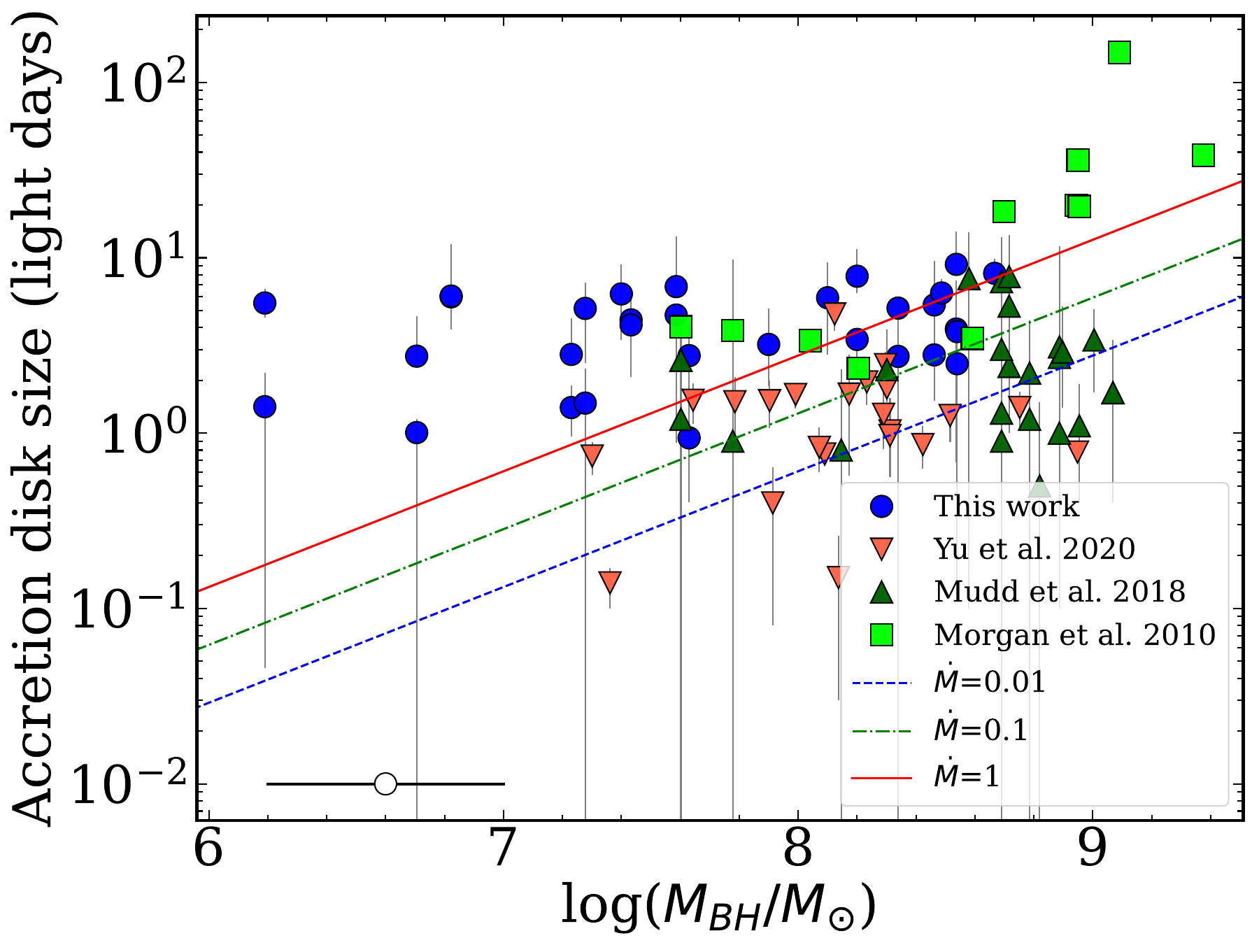}

\caption{ The scaling of the disk sizes with the SMBH masses are shown here. The disk size used here is the effective size of the accretion disk at the wavelength of 2500 {}\AA{} in the units of light days. The sample used in this study is denoted as blue circles, the sources from \citet{Yu2018} are shown as red inverted triangles, sources from \citet{Mudd2018} are shown as green triangles, and the sources taken from \citet{Morgan2010} are shown as green squares. The uncertainty in SMBH mass estimate is approximately 0.4 dex, which is shown as a black line on the bottom left. The predictions for disk size based on the dimensionless mass accretion rates of 0.01 (blue dashed line), 0.1 (green dotted line), and 1 (solid red line) are also shown in the figure.}
\label{Figure8}
\end{figure}

\subsection{ Scaling of Disk size with physical parameters}

We used the disk sizes obtained using the {\sc javelin}  thin disk model to study the relation of disk sizes with the physical parameters, namely the SMBH mass and the luminosity at 5100 \AA{}. According to predictions of the SS disk model, the disk size scales with the SMBH mass as $M^{2/3}$ \citep[see][]{Morgan2010}. The RM-based SMBH mass estimates have been the most robust so far in the absence of any other reliable techniques for SMBH, covering a significant range of redshifts and luminosities. In Figure \ref{Figure8}, we show the relation between the SMBH masses and accretion disk sizes obtained using the {\sc javelin}  thin disk model. We include the previous microlensing-based results from \citet{Morgan2010}. We converted the units of the accretion disk size to light days and converted the reference wavelength at 2500\AA{}. We also included the results from \citet{Mudd2018} and \citet{Yu2018} in this relation. We also plot the expected relation between the SMBH mass and the accretion disk size based on the dimensionless accretion rates of 0.01, 0.1, and 1. The dimensionless accretion rate is  $\dot{M}/\dot{M_{\textsc{edd}}}$, where $\dot{M}$ is the mass accretion rate defined as  $L_{\textsc{bol}}/ \eta c^2$ and $\dot{M_{\textsc{edd}}}$ is the Eddington mass accretion rate defined as $L_{\textsc{edd}}/c^2$ and $L_{ \textsc{edd}}= 1.45\times 10^{38}\times \frac{M_{BH}}{M\odot} erg/s$. $L_{\textsc{bol}}$ is the bolometric luminosituy calculated as roughly 9 times the luminosity at 5100\AA{} \citep{Kaspi2002}. The radiative efficiency is denoted as $\eta$, which we set as 0.1 based on recent works.

We find a weak correlation between the obtained disk sizes and the SMBH masses, with the Spearman rank correlation coefficient being 0.38. Although \citet{Morgan2010} found a relation linking the SMBH mass and disk sizes obtained using microlensing, their sample covered the higher end of SMBH masses only, while our sample covers a range of log (M$_{BH}/M_{\odot}$) ranging from 6.19 to 8.66 which is larger than the range covered in that work. Although the correlation is not very strong, the general trend of AGN with higher SMBH mass tends to have larger accretion disk sizes. Both the disk size and SMBH mass estimates suffer large uncertainties, and thus, proper monitoring and the robust estimates of the accretion disk size will be very helpful in establishing this relation further.

\begin{figure}

\includegraphics[width=8.5cm,height=6cm]{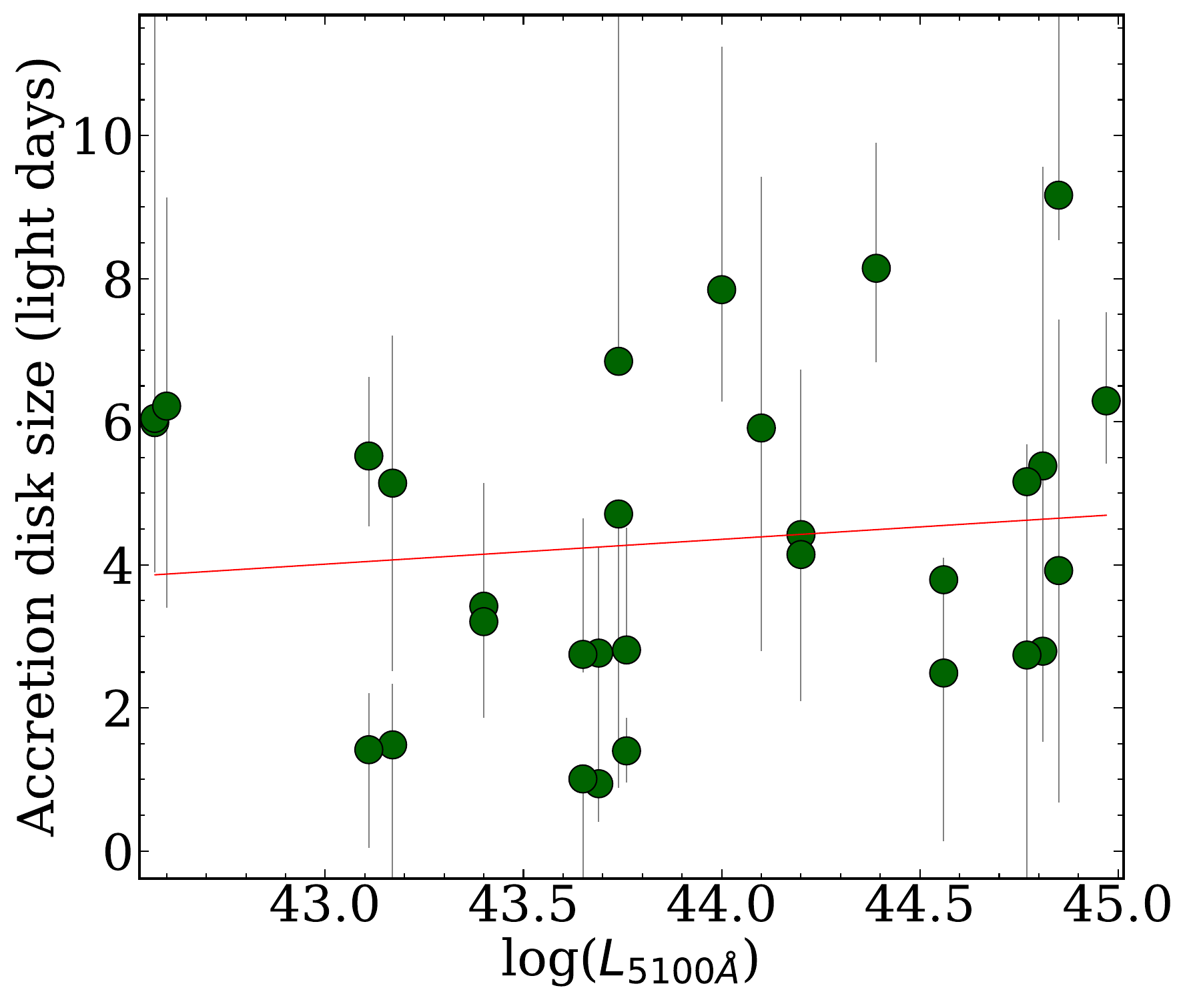}

\caption{ The scaling of the disk sizes corresponding to the  $g$-band rest wavelength with the luminosity at 5100 \AA{} for 19 AGN from this sample. The red line denotes the fitted straight line.}
\label{Figure9}
\end{figure}

Further, in Figure \ref{Figure9}, we show the relation between the accretion disk sizes and the luminosity at 5100\AA{}. We expect that the disk sizes should be related with intrinsic luminosity as more luminous quasars are expected to have large accretion disks and hence larges inter-band lags, based on the light travel times \citep{Jiang2017}. In \citet{Sergeev2014}, the interband lags are found to scale with the luminosity as $\tau=L^{0.4-0.5}$. However, we find a very weak correlation between the luminosities and the disk size, with a Spearman rank correlation coefficient of 0.18 and a null hypothesis ($p_{null}$) value of 0.31.

\section{Discussion}
\label{section6}

To constrain the size of the accretion disk for a sample of quasars with known SMBH masses, we estimated the inter-band reverberation lags between the optical $g$, $r$, and $i$ bands observations from the ZTF survey. The reverberation lags represent the light travel time across the two different regions of the disk. We found that the interband light curves are correlated, and for 19 sources, the lags increasing with wavelength were recovered successfully using the {\sc javelin}  and ICCF methods. The $g-r$ inter-band lags were shorter than the $g-i$ inter-band lags for these sources, which is consistent with the disk reprocessing {\it lamppost} model, according to which the photons arising from the innermost regions are reprocessed in the form of emission from the outer regions resulting in a lag. In multi-wavelength monitoring campaigns involving X-ray, UV, and optical wavebands, such as the AGN STORM campaign \citep{Starkey}, inter-band lags have been reported in all the wavelengths, with longer wavelengths lagging the short wavelengths.
Further, even using only optical band observations, inter-band lags have been successfully recovered for a large sample of AGN \citep[see][]{Jiang2017, Mudd2018, Homayouni2020, Yu2018}. Our results complement these important observations.

We find out that the size of the accretion disk obtained at a reference wavelength is larger than predicted by the Shakura Sunyaev accretion disk model for a majority of the sources (see Figure \ref{Figure7}). We note that the interband lags for 5 sources, namely PG0026+129, NGC 4253, PG1411+442, RMID300, and PG2130+099, follow the predictions of the SS disk model, while 14 sources have lags of about 3 to 4 times larger than the expectations from the SS disk model. However, we could not zero in on a parameter that distinguished these sources from the sources where the disk sizes are much larger than the SS disk assumption. In previous works, the disk size has been known to scale with the wavelength as a power law of index 4/3, but the sizes have been reported to be larger than predicted by the SS disk \citep[][and references therein]{Fausnaugh2016, Jiang2017, Fausnaugh2018}. Contrarily, \citet{Mudd2018} reported that the disk sizes for their sample of 15 sources did not differ much from the SS disk analytical model. Interestingly the disk sizes obtained through microlensing have also reached similar conclusions \citep{Pooley2007, Blackburne2011, Mosquera2013}. Larger accretion disk sizes obtained through both the continuum reverberation mapping and the microlensing methods for a majority of the sources may indicate that the standard SS disk assumption does not hold for the AGN in general and additional components may be needed while modeling the accretion disks in AGN.
We note that some of the sources in our sample have higher accretion rates \citep{Du2015}. It is possible that the accretion mechanism in the AGN with higher accretion rates is different and thus, the disk sizes obtained assuming the SS disk model may not hold true for these kinds of AGN. Evidence of non-disk component in the optical continuum for Mrk 279 has been reported by \citet{Chelouche2019}. This might be a possibility for the larger than expected lags for some of the sources in this sample too. However, the lag measurements suffer large uncertainties in the sources where this effect might take place. Also, the exact contribution from such a component can be achieved through either a combination of narrowband filters or simultaneous spectra, which has not been possible in our case. Another possible explanation for the larger disk sizes has been proposed by \citet{Gaskell2017} where the reddening effects could lead to larger than expected accretion disk sizes. However, \citet{Nunez2019} find out that for Mrk  509, this effect is not able to explain the larger disk sizes.

Continuum reverberation mapping gives more robust estimates with multi-wavelength campaigns, as we see in the case of NGC5548 from the AGN STORM campaign. Nevertheless, with optical data points only, we have been able to constrain the disk sizes for a large number of objects. While the exact structure of the AGN accretion disks remains unknown so far, there is overwhelming evidence in favor of disk reprocessing based on confirmed inter-band lags, increasing with the wavelength \citep{Mchardy2014, Fausnaugh2018}. Our results for a set of reverberation mapped AGN differing in SMBH mass and luminosities provide another set of observational evidence for testing the accretion mechanism in a wide range of such objects. While we find a weak correlation between the obtained disk size, the SMBH mass, and the luminosity, this might be due to the limit of observations as we cannot cover a significant wavelength range using the ZTF dataset. Another reason behind the weak correlation could be the wide range of SMBH masses and luminosities covered in the sample. In the future, with the availability of surveys such as the Vera C. Rubin Observatory Legacy Survey of Space and Time (LSST) \citep{lsst2019}, it will be possible to cover a larger sample with much better cadence and hence will be helpful in better constraining these correlations.

\section{Conclusions}
\label{section7}

We obtained the accretion disk size measurements of AGN with previous SMBH mass estimates through reverberation mapping, with a wide range of luminosities, redshifts, and SMBH masses, based on $g$, $r$, and $i$ band observations from the ZTF survey. The high cadence observations from ZTF provide an excellent dataset to constrain interband reverberation lags efficiently. The primary conclusions from this work are as follows:

\begin{enumerate}
    \item The $g$, $r$, and $i$ band light curves are correlated, and the $g-r$ inter-band lags are shorter than $g-i$ inter-band lags for 19 sources in our sample, which provide strong evidence in favor of the disk reprocessing.
    \item The interband lags and the disk sizes obtained using the {\sc javelin}  thin disk model are larger than the ones predicted by the standard SS thin disk model for a majority of sources, which is consistent with the recent findings and raises the question of the usage of the simple SS disk model for AGN accretion disks.
    \item For 5 sources, we obtained the interband lags in agreement with the SS disk predictions. However, no significant parameter distinguishes these AGN from the ones where we found larger disk sizes.
    \item There is a weak correlation found in the SMBH mass versus disk size and the luminosity versus disk size, which may be due to the uncertainties in both the accretion disk size measurements and the SMBH mass measurements, which have been known to suffer uncertainties up to 0.4 dex.
\end{enumerate}

 Although mapping the entire disk profile is only possible with multi-wavelength campaigns such as the AGN-STORM campaign of NGC 5548  \citep{Edelson2015}, public surveys as the ZTF provide us the opportunity to use light curves for a large number of sources and obtain inter-band lags to understand the accretion mechanisms responsible for the interband lags, albeit with a smaller wavelength coverage. This work can be extended with larger AGN samples and derive the accretion mechanisms powering them to understand the AGN accretion disks better.

\section*{Acknowledgements}
We are thankful to the anonymous referee for providing comments and suggestions, which helped us to improve our manuscript. LCH and XBW were supported by the National Science Foundation of China (11721303, 11991052, 12133001) and the National Key R\&D Program of China (2016YFA0400702, 2016YFA0400703). This research is part of the DST-SERB project under grant no. EMR/2016/001723. VKJ and HC acknowledge the financial support provided by DST-SERB for this work. VKJ is thankful to Suvendu Rakshit (ARIES, Nainital) for useful discussion and help regarding the {\sc javelin} code. VKJ acknowledges the warm hospitality provided by CUHP Dharamshala during his stay. Based on observations obtained with the Samuel Oschin 48-inch Telescope at the Palomar Observatory as part of the Zwicky Transient Facility project. ZTF is supported by the National Science Foundation under Grant No. AST-1440341 and collaboration including Caltech, IPAC, the Weizmann Institute for Science, the Oskar Klein Center at Stockholm University, the University of Maryland, the University of Washington, Deutsches Elektronen-Synchrotron and Humboldt University, Los Alamos National Laboratories, the TANGO Consortium of Taiwan, the University of Wisconsin at Milwaukee, and Lawrence Berkeley National Laboratories. Operations are conducted by COO, IPAC, and UW.

\section*{Data Availability}

The data being used for this study is publicly available in the ZTF DR6 database.



\bibliographystyle{mnras}
\bibliography{main} 





\bsp	
\label{lastpage}
\end{document}